\documentclass[aps,prab,preprint,superscriptaddress,endfloats,showpacs]{revtex4}

\usepackage[utf8]{inputenc}
\usepackage{amsmath}
\usepackage{graphicx}
\usepackage{epstopdf}
\usepackage{color}
\usepackage{tabularx}

\begin{document}

\title{Statistical properties of a free-electron laser revealed by the Hanbury Brown and Twiss interferometry}
\date{\today}

\author{O. Yu.~Gorobtsov}
	\affiliation{Deutsches Elektronen-Synchrotron DESY, Notkestra{\ss}e 85, D-22607 Hamburg, Germany}
\author{G.~Mercurio}
	\affiliation{Department of Physics, University of Hamburg and Center for Free Electron Laser Science, Luruper Chausse 149, D-22761 Hamburg, Germany}
\author{G.~Brenner}
\affiliation{Deutsches Elektronen-Synchrotron DESY, Notkestra{\ss}e 85, D-22607 Hamburg, Germany}
\author{U.~Lorenz}
\affiliation{Institute of Chemistry, University of Potsdam, D-14476 Potsdam, Germany}
\author{N.~Gerasimova}
\affiliation{European XFEL GmbH, Holzkoppel 4, 22869 Schenefeld, Germany}
\author{R.P.~Kurta}
\affiliation{European XFEL GmbH, Holzkoppel 4, 22869 Schenefeld, Germany}
\author{F.~Hieke}
\affiliation{Department of Physics, University of Hamburg and Center for Free Electron Laser Science, Luruper Chausse 149, D-22761 Hamburg, Germany}	
\author{P.~Skopintsev}
\affiliation{Deutsches Elektronen-Synchrotron DESY, Notkestra{\ss}e 85, D-22607 Hamburg, Germany}
\affiliation{Moscow Institute of Physics and Technology (State University), Dolgoprudny, 141700 Moscow Region, Russia}
\author{I.~Zaluzhnyy}
\affiliation{Deutsches Elektronen-Synchrotron DESY, Notkestra{\ss}e 85, D-22607 Hamburg, Germany}
\affiliation{National Research Nuclear University MEPhI (Moscow Engineering Physics Institute), Kashirskoe shosse 31, 115409 Moscow, Russia }	
\author{S.~Lazarev}
\affiliation{Deutsches Elektronen-Synchrotron DESY, Notkestra{\ss}e 85, D-22607 Hamburg, Germany}
\affiliation{National Research Tomsk Polytechnic University (TPU), pr. Lenina 2a,
	634028 Tomsk, Russia}
\author{D.~Dzhigaev}
\affiliation{Deutsches Elektronen-Synchrotron DESY, Notkestra{\ss}e 85, D-22607 Hamburg, Germany}
\author{M.~Rose}
\affiliation{Deutsches Elektronen-Synchrotron DESY, Notkestra{\ss}e 85, D-22607 Hamburg, Germany}
\author{A.~Singer}
\affiliation{University of California San Diego, 9500 Gilman Dr., La Jolla, California 92093, United States}
\author{W.~Wurth}
\affiliation{Deutsches Elektronen-Synchrotron DESY, Notkestra{\ss}e 85, D-22607 Hamburg, Germany}
\affiliation{Department of Physics, University of Hamburg and Center for Free Electron Laser Science, Luruper Chausse 149, D-22761 Hamburg, Germany}
\author{I. A. ~Vartanyants}
\email[Corresponding author: ]{ivan.vartaniants@desy.de}
\affiliation{Deutsches Elektronen-Synchrotron DESY, Notkestra{\ss}e 85, D-22607 Hamburg, Germany}
\affiliation{National Research Nuclear University MEPhI (Moscow Engineering Physics Institute), Kashirskoe shosse 31, 115409 Moscow, Russia }

\begin{abstract}

We present a comprehensive experimental analysis of statistical properties of the self-amplified spontaneous emission (SASE) free electron laser (FEL) FLASH by means of Hanbury Brown and Twiss (HBT) interferometry.
The experiments were performed at the FEL wavelengths of 5.5 nm, 13.4 nm, and 20.8 nm.
We determined the 2-nd order intensity correlation function for all wavelengths and different operation conditions of FLASH.
In all experiments a high degree of spatial coherence (above 50\%) was obtained.
Our analysis performed in spatial and spectral domains provided us with the independent measurements of an average pulse duration of the FEL that were below 60 fs.
To explain complicated behavior of the 2-nd order intensity correlation function we developed advanced theoretical model that includes the presence of multiple beams and external positional jitter of the FEL pulses.
By this analysis we determined that in one of the experiments external positional jitter was about 25\% of the beam size.
We envision that methods developed in our study will be used widely for analysis and diagnostics of the FEL radiation.
%

\end{abstract}


\maketitle

\section{Introduction}

X-ray Free Electron Lasers (XFELs) are presently the brightest  sources of x-ray radiation \cite{LCLS,SCSS,XFEL}.
They find applications in a wide range of fields: structural biology \cite{Seibert2011, Chapman2011}, solid density plasma \cite{Vinko2012}, studies of local properties and dynamics of matter under extreme conditions \cite{Schropp2015}, ultrafast photochemistry \cite{LiekhusSchmaltz2015}, atomic physics \cite{Young2010}, condensed matter physics \cite{Trigo2013, Dronyak2012, Clark2013, Singer2016, Beaud2014} and many others.
The unique combination of femtosecond pulse duration, high brilliance and degree of coherence advances the state of the art in conventional x-ray methods and gives access to new techniques that were not feasible with previous x-ray facilities.
High degree of spatial coherence in comparison with other x-ray sources is the key property of the XFELs.
X-ray scattering with highly coherent illumination opens the possibility for high resolution coherent diffraction imaging for a wide range of systems \cite{Nugent2010, Chapman2010, Vartanyants2015}.
However, insufficient coherence leads to decreased contrast in coherent imaging technique, reduces the reconstructed image resolution and, in the worst case, makes it impossible to retrieve the desired structural information \cite{Vartanyants2001, Vartanyants2003, Williams2007, Whitehead2009}.

From that respect a deep analysis and characterization of statistical properties of FELs is an extremely important task.
It is known that self-amplified spontaneous emission (SASE) FELs, for which the amplification process starts from shot noise of the electron bunch, radiate as chaotic sources with high spatial coherence \cite{SaldinBook, Saldin2006}.
Longitudinal coherence of SASE FEL radiation, which is slowly growing with the undulator length, is typically much lower in comparison to spatial one \cite{Saldin2006}.
A large variety of methods can be used to determine transverse and longitudinal coherence properties of x-rays, such as measurements with double pinholes or arrays of slits \cite{Vartanyants2011, Singer2012, Skopintsev2014}, Michelson type interferometry \cite{Roling2011, Singer2012, Hilbert2014} or Hanbury Brown and Twiss (HBT) intensity interferometry \cite{Singer2013} (see for review \cite{Vartaniants2015a}).
Among these methods, HBT interferometry provides additional information on the second- and higher order statistical properties of FEL sources, while removing the need for additional optical devices.

Intensity interferometry, as introduced by Hanbury Brown and Twiss \cite{Hanbury, Twiss}, was a revolutionary experiment at the time.
Their measurements, seemingly showing contradiction between classical and quantum theories of light, led to the development of quantum optics \cite{Glauber1963}.
Since then, HBT interferometry has found applications in many areas of physics.
For example, it was used to probe Bose-Einstein condensates \cite{Schellekens2005} and to analyze nuclear scattering experiments \cite{Baym1998}.
Intensity-intensity correlation measurements in x-ray energy range were first suggested to solve the phase problem in crystallography \cite{Goldberger}, further developed conceptually \cite{Gluskin1991, Ikonen1992} and finally performed at synchrotron sources \cite{Gluskin1999, Yabashi2001, Yabashi2002, Singer2014}.
Correlation of intensities at two points in space, expressed in terms of the degree of second-order coherence, is particularly well suited for interferometry at FEL sources \cite{Singer2013, Song2014}.
The femtosecond pulse duration of the FEL radiation allows to eliminate the necessity for a correlator device to perform coincidence measurements.
In comparison with Young's interferometry, where the double pinholes separation must be changed to measure coherence between different spatial positions \cite{Vartanyants2011, Singer2012}, the HBT approach with a pixelated detector allows to measure correlation function at a set of distances simultaneously.
As soon as intensities and not amplitude correlations are measured in an HBT experiment, it is not sensitive to phase fluctuations, which can significantly affect Young's or Michelson interferometry \cite{Goodman, MandelWolf}.
Importantly, HBT provides a possibility of high-order statistical analysis of FEL radiation properties.

Here, we present a comprehensive analysis of the statistical properties of FLASH FEL radiation in non-linear regime of operation.
It is based on HBT interferometry experiments performed at different FEL wavelengths and radiation parameters.
Conventional HBT theory assuming a single distant incoherent source does not explain our experimental results.
We develop advanced theoretical approach that includes effects of multiple secondary beams and external positional jitter to explain observed experimental features.
These effects can be revealed only through the intensity-intensity interferometry.

\section{Theoretical principles of the intensity-intensity interferometry}


The basic theory of optical coherence is the second-order coherence theory, when amplitudes of the wavefields are correlated \cite{Goodman, MandelWolf}.
It provides an appropriate treatment for most of the traditional interference phenomena, such as Young's double pinhole or Michelson type of experiments.
However, for a full understanding of certain experimental results, the basic coherence theory is not sufficient.
For instance, the statistical difference between a laser and coherent chaotic source can only be revealed through a higher-order coherence theory \cite{MandelWolf}.
In this work, we are mostly interested in the second order intensity correlations, which will be now introduced through a more general concept of the fourth order amplitude correlations.

Assuming a scalar field $E(\mathbf{r},t)$, the cross-correlation function of the fourth order can be defined as \cite{MandelWolf}
\begin{equation}
\Gamma^{(4)}(\mathbf{r_1}, t_1; ...;\mathbf{r_{4}}, t_{4}) = \langle E^*(\mathbf{r_1}, t_1) E^*(\mathbf{r_2}, t_2) E(\mathbf{r_{3}}, t_{3}) E(\mathbf{r_{4}}, t_{4}) \rangle\ .
\label{coherence::4th_order_correlation}
\end{equation}
where the angle brackets denote the ensemble average, and $\mathbf{r}, t$ are space and time coordinates, respectively.

\begin{figure}
	\includegraphics[width=\linewidth]{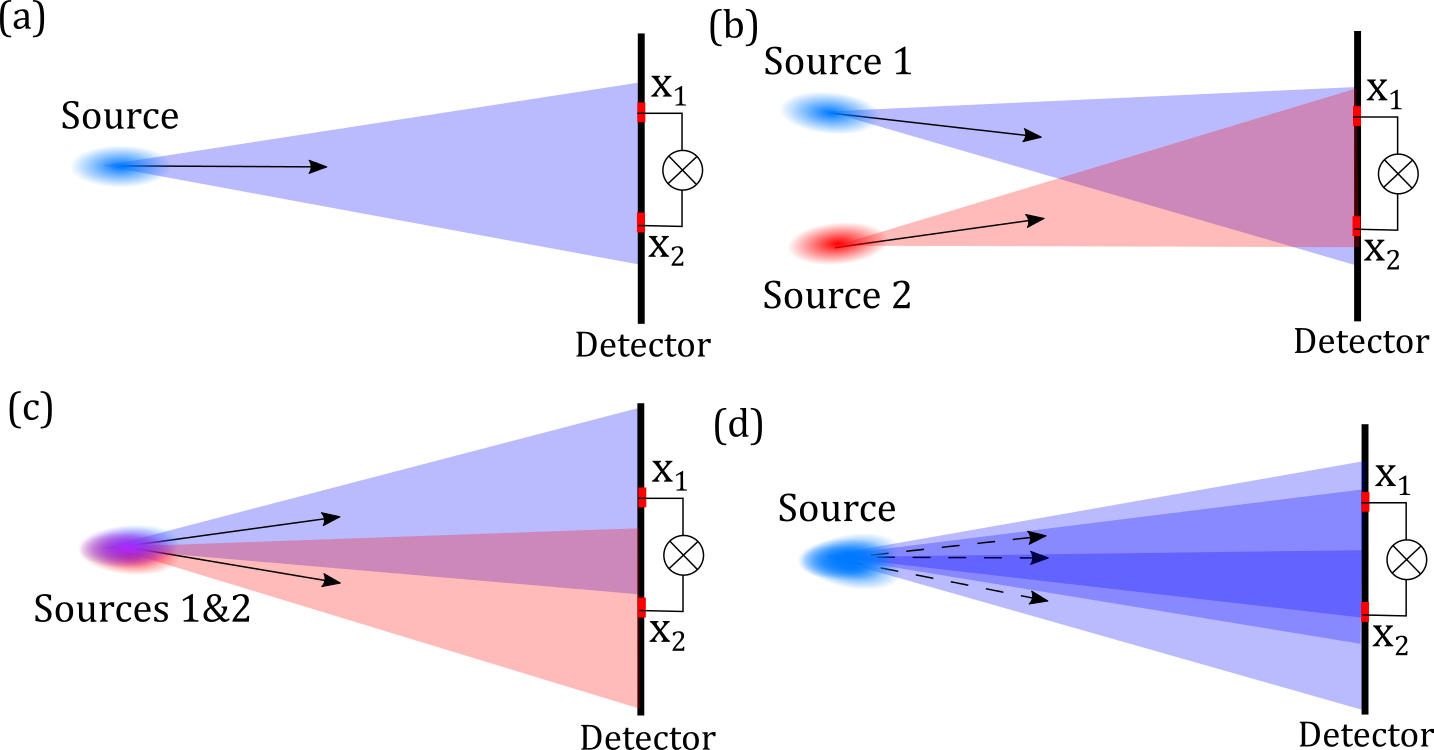}
	\caption{\label{fig::Different_sources}
            Different configuration of the sources. All sources are chaotic extended sources at large distance from the detector. Intensity-intensity correlation is measured as a coincidence
            signal at two pixels of the detector
        (a) Single source.
		(b) Two sources separated from each other.
		(c) Single source radiating in multiple angular positions that results in multiple beam illumination of the detector.
        (d) Single source with the angular jitter that results in an extended illumination of the detector.
	}
	\newpage
\end{figure}

In an HBT experiment, the intensities incoming from a distant source (or several sources) are measured simultaneously at different spatial positions (see Fig.~\ref{fig::Different_sources}).
The measured intensity is an integral of the instantaneous intensity $I(\mathbf{r},t) = E^*(\mathbf{r},t)E(\mathbf{r},t)$ over the detector time resolution.
When HBT interferometry is performed for stationary and ergodic sources such as stars \cite{Hanbury}, the repeated measurements within a limited time window can be used to calculate the ensemble average.
At a free-electron laser, which is a pulsed and therefore a non-stationary source, the averaging is performed over different pulses \footnote{We assume here that all pulses are realizations of the same statistical process.}.
Since the FEL pulse duration is much shorter than the detector currently availible time resolution, we naturally get time integrated values of intensity
\begin{equation}
I(\mathbf r)=\int\limits_{-\infty}^{\infty} I(\mathbf{r},t)\mbox dt = \int\limits_{-\infty}^{\infty} |E(\mathbf r,t)|^2\mbox dt\ ,
\label{coherence::intensity_space_time}
\end{equation}
or in the space-frequency domain, defining $E(\mathbf r,t) = \int E(\mathbf r,\omega) \exp(i\omega t)\mbox d\omega/(2\pi)$, where $E(\mathbf r,\omega)$ is the Fourier spectral component of the field $E(\mathbf r,t)$, we obtain
\begin{equation}
I(\mathbf r) = \int\limits_{-\infty}^{\infty} |E(\mathbf r,\omega)|^2\frac{\mbox d\omega}{2\pi}\ .
\label{coherence::intensity_space_frequency}
\end{equation}
In the case of radiation with a limited bandwidth, defined, for example, by a monochromator, we can formally introduce the complex transmittance function $T(\omega)$ and have for the intensity
\begin{equation}
I(\mathbf r)=\int\limits_{-\infty}^{\infty} |T(\omega)|^2 |E(\mathbf r,\omega)|^2\frac{\mbox d\omega}{2\pi}\ .
\label{coherence::monochromator_transmits}
\end{equation}
The second-order intensity correlation function can be now defined as
\begin{equation}
g^{(2)}(\mathbf{r_1}, \mathbf{r_{2}}) = \frac{\langle I(\mathbf{r_1}) I(\mathbf{r_2}) \rangle}{\langle I(\mathbf{r_1})\rangle \langle I(\mathbf{r_2}) \rangle}\ .
\label{coherence::2_order_normalized}
\end{equation}
%

We introduce also a spectral cross-correlation function in the space-frequency domain \cite{MandelWolf}.
Since we are only interested in intensity correlations, we will further use the following definition for the fourth-order spectral cross-correlation function
$W^{(4)}(\mathbf{r_1}, \omega_1; \mathbf{r_2}, \omega_2) \equiv \langle E^{\mbox*}(\mathbf{r_1}, \omega_1) E(\mathbf{r_1}, \omega_1 ) E^{\mbox*}(\mathbf{r_2}, \omega_2) E(\mathbf{r_2}, \omega_2 )\rangle$.
Now substituting Eq. \eqref{coherence::monochromator_transmits} into $g^{(2)}(\mathbf{r}_1,\mathbf{r}_2)$ defined by Eq. \eqref{coherence::2_order_normalized} we obtain
\begin{equation}
g^{(2)}(\mathbf{r}_1,\mathbf{r}_2)
= \frac{\iint\limits_{-\infty}^{\infty} |T(\omega_1)|^2|T(\omega_2)|^2 W^{(4)}(\mathbf{r_1},\omega_1;\mathbf{r_2},\omega_2)\mbox d\omega_1 \mbox d\omega_2}
{\int\limits_{-\infty}^{\infty}|T(\omega_1)|^2 S(\mathbf{r_1},\omega_1)\mbox d\omega_1\int\limits_{-\infty}^{\infty}|T(\omega_2)|^2 S(\mathbf{r_2},\omega_2) \mbox d\omega_2} \ .
\label{coherence::g2_general}
\end{equation}

Here we introduced the spectral density $S(\mathbf{r}, \omega) = W^{(2)}(\mathbf{r}, \omega; \mathbf{r}, \omega)$, where the second-order spectral cross-correlation function is defined as $W^{(2)}(\mathbf{r_1}, \omega_1; \mathbf{r_2}, \omega_2) = \langle E^*(\mathbf{r_1}, \omega_1) E(\mathbf{r_2}, \omega_2 ) \rangle $.

First, we will describe the behavior of the $g^{(2)}(\mathbf{r_1}, \mathbf{r_{2}})$ function for a chaotic source.
Radiation properties of such a source can be sufficiently described in the framework of Gaussian statistics \cite{MandelWolf}.
Applying the Gaussian moment theorem \cite{MandelWolf} we obtain for
$W^{(4)}(\mathbf{r_1},\omega_1; \mathbf{r_2}, \omega_2) = S(\mathbf{r_1}, \omega_1)S(\mathbf{r_2}, \omega_2)+|W^{(2)}(\mathbf{r_1}, \omega_1; \mathbf{r_2}, \omega_2)|^2$. Substituting this relation into Eq. \eqref{coherence::g2_general} we can derive the expression for the second-order intensity correlation function of the Gaussian source as
\begin{equation}
g^{(2)}(\mathbf{r}_1,\mathbf{r}_2)
= 1+\frac{\iint\limits_{-\infty}^{\infty} |T(\omega_1)|^2|T(\omega_2)|^2 |W^{(2)}(\mathbf{r_1},\omega_1,\mathbf{r_2},\omega_2)|^2\mbox d\omega_1 \mbox d\omega_2}
{\int\limits_{-\infty}^{\infty}|T(\omega)|^2 S(\mathbf{r_1},\omega)\mbox d\omega\int\limits_{-\infty}^{\infty}|T(\omega)|^2 S(\mathbf{r_2},\omega) \mbox d\omega}\ .
\label{coherence::g2_entangled}
\end{equation}

In the following we will consider the radiation to be cross-spectrally pure \cite{MandelWolf,Goodman}.
For such radiation the spectral cross-correlation function is separable into its spatial and spectral components: $W^{(2)}(\mathbf{r_1},\mathbf{r_2};\omega_1,\omega_2)=J(\mathbf{r_1}, \mathbf{r_2})W(\omega_1, \omega_2)$ and $S(\mathbf{r},\omega)=I(\mathbf{r})S(\omega)$, where $J(\mathbf{r_1}, \mathbf{r_2})=\langle E^*(\mathbf{r_1}) E(\mathbf{r_2}) \rangle$ is the mutual intensity (in this case intensity $I(\mathbf{r})=J(\mathbf{r},\mathbf{r})$) and $W(\omega_1, \omega_2)=\langle E^*(\omega_1) E(\omega_2) \rangle$.
In this case the second-order correlation function defined in Eq. \eqref{coherence::g2_entangled} can be expressed as \cite{Singer2013}
\begin{equation}
g^{(2)}(\mathbf{r_1}, \mathbf{r_{2}}) = 1 + \zeta_2(D_{\omega}) |\mu(\mathbf{r_1},\mathbf{r_{2}})|^2\ .
\label{coherence::g2_traditional}
\end{equation}
Here $\zeta_2(D_{\omega})$ is the contrast function that strongly depends on the radiation frequency bandwidth $D_{\omega}$ and is defined as
\begin{equation}
\zeta_2(D_{\omega}) = \frac{\iint\limits_{-\infty}^{\infty} |T(\omega_1)|^2|T(\omega_2)|^2|W(\omega_1,\omega_2)|^2 \mbox d\omega_1 \mbox d\omega_2}
{\Bigl(\int\limits_{-\infty}^{\infty}|T(\omega)|^2S(\omega)\mbox d\omega \Bigr)^2}\ ,
\label{coherence::zeta_2}
\end{equation}
and $\mu(\mathbf{r_1}, \mathbf{r_2})$ is the normalized spectral degree of coherence \cite{Goodman}
\begin{equation}
\mu(\mathbf{r_1}, \mathbf{r_2}) = \frac{J(\mathbf{r_1},\mathbf{r_2})}{\sqrt{I(\mathbf{r_1})I(\mathbf{r_2})}} \ .
\label{coherence::mu2}
\end{equation}

It can be shown \cite{Ikonen1992} that in the case of a stationary chaotic beam $\zeta_2(D_{\omega})$ is determined by $\tau_c/T$ , where $\tau_c = 2\pi/D_{\omega}$ is the coherence time and $T$ is the pulse duration of the FEL radiation.
In this limit the number of longitudinal modes $M_t$ that is defined as $M_t = T/\tau_c$ is inversely proportional to the contrast function $\zeta_2(D_{\omega})$.
Notice also that for $\mathbf{r_1}=\mathbf{r_2}=\mathbf{r}$ $g^{(2)}(\mathbf{r}, \mathbf{r}) = 1+\zeta_2(D_{\omega})$ and does not depend on position $\mathbf{r}$.

To characterize the global spatial coherence properties of the FEL beam we also introduce the spatial degree of transverse coherence $\zeta_s$ as \cite{Saldin2008, Vartanyants2010}
\begin{equation}
\zeta_s = \frac{\int|J(\mathbf{r_1},\mathbf{r_2})|^2 d\mathbf{r_1} d\mathbf{r_2}}{\left(\int I(\mathbf{r}) d\mathbf{r} \right)^2} =  \frac{\int|\mu(\mathbf{r_1},\mathbf{r_2})|^2 I(\mathbf{r_1})I(\mathbf{r_2})d\mathbf{r_1} d\mathbf{r_2}}{\left(\int I(\mathbf{r}) d\mathbf{r} \right)^2} \ .
\label{coherence::coherence_degree_definition}
\end{equation}
Using expression \eqref{coherence::g2_traditional} we can determine degree of spatial coherence through the 2-nd order intensity correlation function as
\begin{equation}
\zeta_s = \frac{1}{\zeta_2(D_{\omega})}\frac{\int \left[g^{(2)}(\mathbf{r_1}, \mathbf{r_{2}})-1\right] I(\mathbf{r_1})I(\mathbf{r_2})d\mathbf{r_1} d\mathbf{r_2}}{\left(\int I(\mathbf{r}) d\mathbf{r} \right)^2}\ .
\label{coherence::coherence_degree_expressed}
\end{equation}

To describe FEL statistical pulse properties with a simple model we considered that the FEL beam spectral cross-correlation function can be characterized in the framework of the Gaussian Schell-model (GSM) \cite{Lajunen2005}
\begin{equation}
W^{(2)}(x_1, \omega_1; x_2,\omega_2) = W_{0} J(x_1,x_2)W(\omega_1,\omega_2)\ ,
\label{coherence::GSM}
\end{equation}
where $W_{0}$ is the normalization constant and
\begin{equation}
        \begin{aligned}
        J(x_1,x_2) & = &
\exp\left[-\frac{(x_1-x_0)^2+(x_2-x_0)^2}{4\sigma_x^2}-\frac{(x_1-x_2)^2}{2\xi_x^2}\right]\
, \\
        W(\omega_1,\omega_2) & = &
\exp\left[-\frac{(\omega_1-\omega_0)^2-(\omega_2-\omega_0)^2}{4\Omega^2}-\frac{(\omega_1-\omega_2)^2}{2\Omega_c^2}\right]\
.
        \end{aligned}
        \label{coherence::GSM_JW}
\end{equation}
Here $x_0$ is the position of the pulse center, $\sigma_x$ is the beam size,  $\xi_x$ is the transverse coherence length, $\omega_0$ is the central pulse frequency, $\Omega$ is the spectral width, and $\Omega_c$ is the spectral coherence width.

We further assume that the transmission function $T(\omega)$ is described by a rectangular function
\begin{equation}
T(\omega) = \left\{ \begin{array}{ll}
1 & \mbox{if $|\omega| \leq D_{\omega}/2$}\ ; \\
0 & \mbox{if $|\omega| > D_{\omega}/2$}\ .\end{array}\right.
\label{coherence::Tomega}
\end{equation}
Substituting Eq. \eqref{coherence::Tomega} into Eq. \eqref{coherence::zeta_2} with the assumption that the monochromator bandwidth is much smaller than the original pulse bandwidth, we obtain for the contrast function \cite{Goodman, Singer2013}
\begin{equation}
\zeta_2(D_{\omega}) = \frac{\sqrt{\pi}}{D_{\omega}T}\text{erf}(D_{\omega}T) + \frac{1}{(D_{\omega}T)^2}(e^{-(D_{\omega}T)^2}-1)\ ,
\label{coherence::contrast_duration_fit}
\end{equation}
where $\text{erf}(x)$ is the error function.
In the limit $D_{\omega}T\rightarrow 0$ the function $\zeta_2(D_{\omega})$ approaches unity asymptotically as $1-(D_{\omega}T)^2/6$ and
in the limit $D_{\omega}T\rightarrow \infty$ it goes to zero as $\sqrt{\pi}/(D_{\omega}T)\sim \tau_c/T$.

\begin{figure}
	\includegraphics[width=\linewidth]{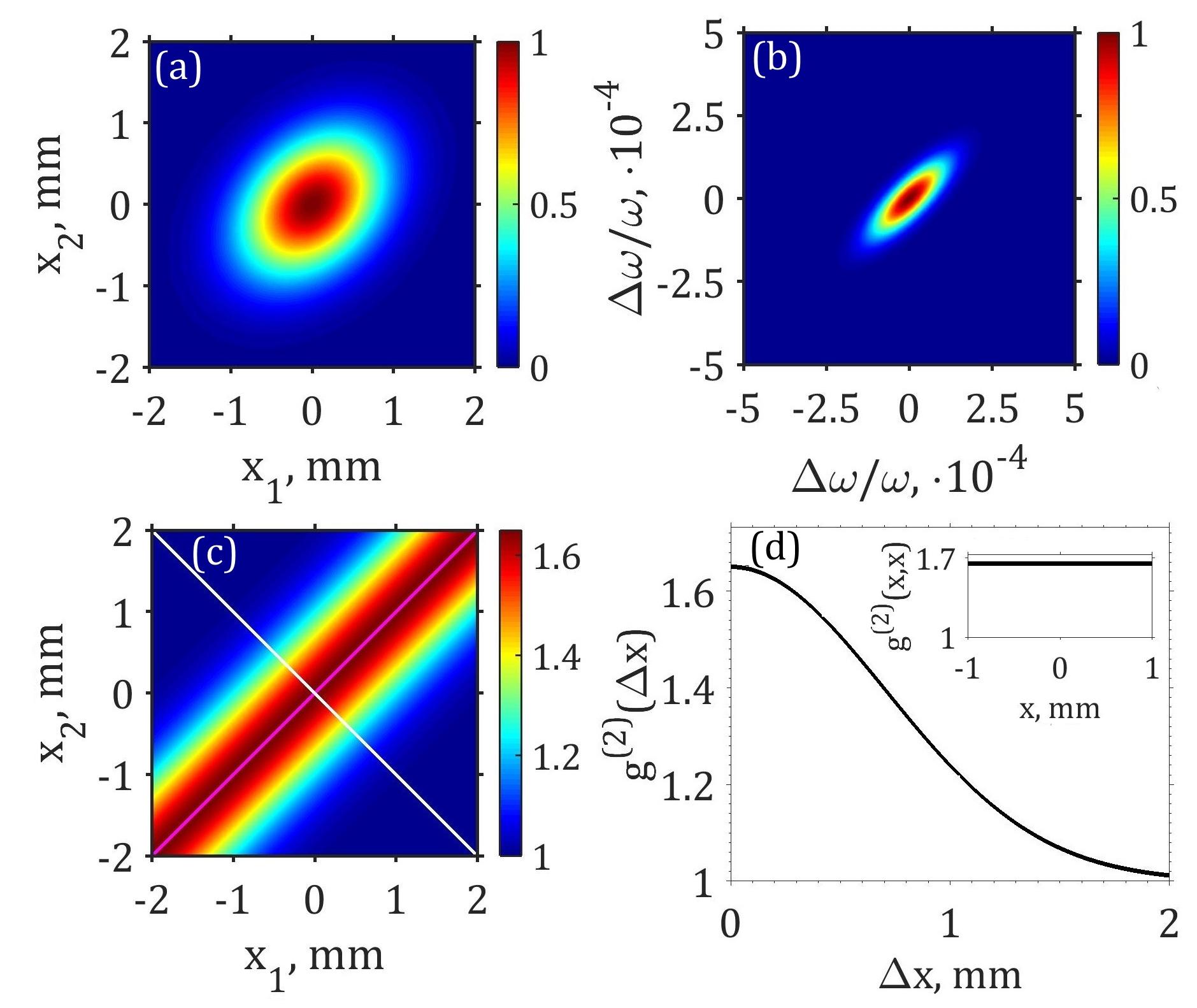}
	\caption{\label{fig::One_source}
		Simulations of the intensity-intensity correlation function for a single partially coherent GSM beam (see Fig.~\ref{fig::Different_sources}(a)) with parameters of the beam listed in Table~\ref{tab::sim_param} as Model I.
        (a) Spatial part of the spectral cross-correlation function $J(x_1,x_2)$.
		(b) Spectral part of the spectral cross-correlation function $W(\omega_1,\omega_2)$.
		(c) Intensity-intensity correlation function $g^{(2)}(x_1,x_2)$.
		(d) Intensity-intensity correlation function values $g^{(2)}(\Delta x)$ taken along the white lines in (c); (inset) autocorrelation function $g^{(2)}(x,x)$ taken along the
            purple line in (c).
	}
	\newpage
\end{figure}

We will consider now a single extended distant source (see Fig.~\ref{fig::Different_sources}(a)) in the frame of GSM described by Eqs.~(\ref{coherence::GSM}-\ref{coherence::GSM_JW}) and parameters listed in Table~\ref{tab::sim_param} as "Model I".
Radiation parameters were chosen to be close to FLASH experimental parameters described further.
In Fig. \ref{fig::One_source} (a, b) cross correlation functions in spatial and spectral domains $J(x_1,x_2)$ and $W(\omega_1,\omega_2)$ are shown, respectively.
The intensity-intensity correlation function $g^{(2)}(x_1,x_2)$ for such a beam is shown in Fig. \ref{fig::One_source} (c).
Its cross-section $g^{(2)}(x_0-\Delta x/2,x_0+\Delta x/2)$ taken along the white line is connected according to Eq.~\eqref{coherence::g2_traditional} to the square modulus of the spectral degree of coherence $|\mu(x_0-\Delta x/2,x_0+\Delta x/2)|^2$ and is shown in Fig. \ref{fig::One_source} (d).
It has a Gaussian shape with the given value of the coherence length $\xi_x = 1$ mm.
Here as for any cross-spectrally pure chaotic source the function $g^{(2)}(x_1,x_2)$ is uniform along the main diagonal $x_1=x_2=x$ which we will also call autocorrelation direction (see Fig. \ref{fig::One_source} (d), inset).

\begin{figure}
	\includegraphics[width=\linewidth]{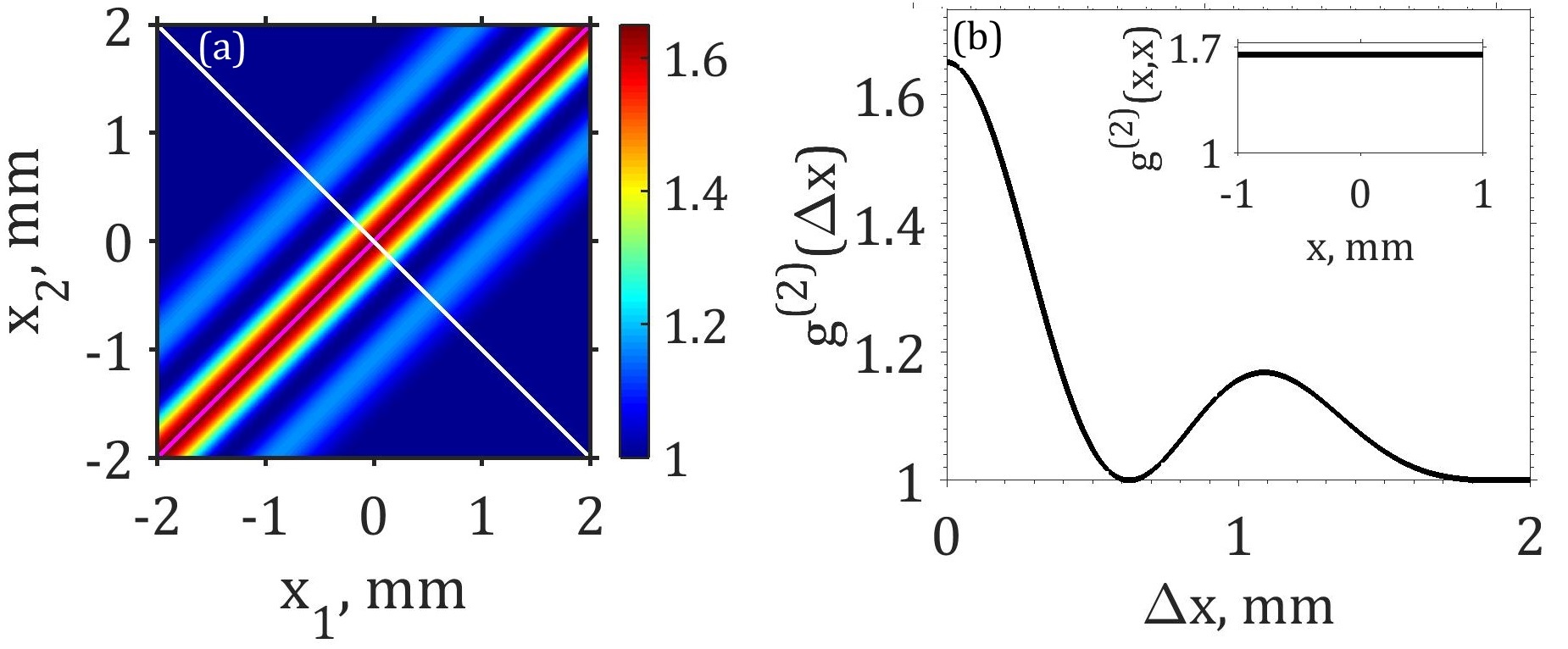}
	\caption{\label{fig::Two_angled}
		Simulations of the intensity-intensity correlation function for two identical partially coherent GSM beams separated from each other (see Fig.~\ref{fig::Different_sources}(b)) with parameters of the beams listed in Table~\ref{tab::sim_param} as Model I.
		(a) Intensity-intensity correlation function $g^{(2)}(x_1,x_2)$.
		(b) Intensity-intensity correlation function values $g^{(2)}(\Delta x)$ taken along the white lines in (a); (inset) autocorrelation function $g^{(2)}(x,x)$ taken along the
            purple line in (a).
		}
	\newpage
\end{figure}

We will consider also another case often discussed in the literature \cite{Baym1998, Paul1986}, that is the case of two distant and independent chaotic sources separated from each other (see Fig.~\ref{fig::Different_sources}(b)).
Such configuration of the sources can be also represented by two independent GSM beams with a relative linear phase shift to each other
\begin{equation}
\begin{aligned}
W^{(2)}(x_1, \omega_1; x_2,\omega_2) = W_1^{(2)}(x_1, \omega_1;
x_2,\omega_2)+W_2^{(2)}(x_1, \omega_1; x_2,\omega_2) \ , \\
J_2(x_1,x_2) = J_1(x_1,x_2) e^{i\Delta k_x(x_2-x_1)} \ , \
W_2(\omega_1,\omega_2) = W_1(\omega_1,\omega_2) \ ,
\end{aligned}
\label{coherence::phaseshift_2beams}
\end{equation}
where  $\Delta k_x$ is the magnitude of the phase shift and $\Delta k_x/k_x$ is the angular separation between these two sources in small angle approximation.
Substituting Eq. \eqref{coherence::phaseshift_2beams} into Eq. \eqref{coherence::g2_entangled} and taking into consideration definition of the contrast function $\zeta_2(D_{\omega})$ in Eq.~\eqref{coherence::zeta_2} we obtain for the intensity-intensity correlation function
\begin{equation}
g^{(2)}(x_1, x_2) = 1 + \zeta_2(D_{\omega}) |\mu(x_1,x_2)|^2 \cos^2\left[ \frac{\Delta k_x(x_2-x_1)}{2} \right] \ .
\label{coherence::g2_traditional_2beams}
\end{equation}
In Fig. \ref{fig::Two_angled}(a) the 2-nd order intensity correlation function  $g^{(2)}(x_1,x_2)$ calculated according to Eq. \eqref{coherence::g2_traditional_2beams} for two beams with the same parameters as in Fig. \ref{fig::One_source} and $\Delta k_x = 5$ mm$^{-1}$ is shown.
The intensity correlation function $g^{(2)}(x_0-\Delta x/2,x_0+\Delta x/2)$ in this case has a characteristic oscillatory behavior (see Fig.~\ref{fig::Two_angled} (b)).
It is important to note that the oscillation of $g^{(2)}(x_0-\Delta x/2,x_0+\Delta x/2)$ is determined mainly by the angular separation of two sources.
While additional structure is clearly visible in $g^{(2)}(x_1, x_2)$, the normalized autocorrelation function $g^{(2)}(x, x)$ is still constant, as demonstrated in Fig. \ref{fig::Two_angled} (b), inset.

\section{Experimental conditions}

\begin{figure}
	\includegraphics[width=\linewidth]{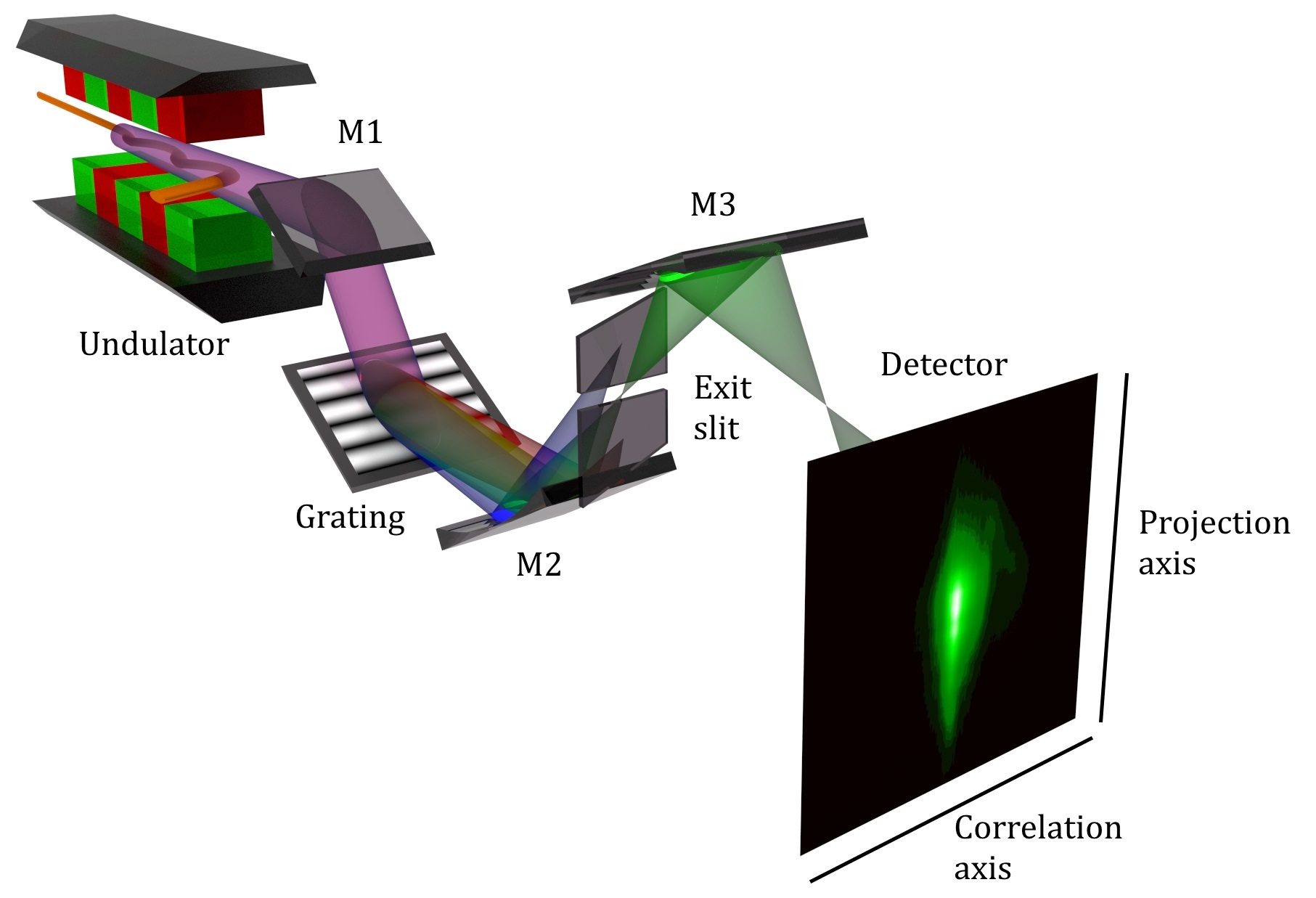}
	\caption{\label{fig::Scheme}
		Schematic layout of the experiment.
        FEL radiation from the undulator is collimated at the mirror M1 and diffracted at the grating.
        The monochromatized radiation is further transmitted through the focusing mirrors M2 and M3 and the exit slit system.
        Intensity profiles are measured at the detector positioned downstream from the focal plane.
	}
	\newpage
\end{figure}

The experiments were performed at FLASH at DESY in deep saturation regime in the XUV energy range at three wavelengths: 5.5 nm, 13.4 nm and 20.8 nm (see Table~\ref{tab::parameters}).
Results of the measurements at 5.5 nm were reported before \cite{Singer2013} and are presented here for completeness.
All measurements were carried out at PG2 beamline \cite{Martins2006, Gerasimova2011}.
Schematic representation of the beamline is shown in Fig.~\ref{fig::Scheme}.
The beam was focused at the position approximately 71.5 m downstream from the undulator exit.
A plane-grating monochromator with a line density of 200 lines/mm was used to vary the bandwidth of the incoming radiation.
Monochromaticity of the incoming beam can be conveniently varied by the size of the monochromator exit slit.
The values of the grating parameter $c_{ff}$, diffraction grating orders, and corresponding photon energy dispersion in the exit slit plane used in the experiment are summarized in Table~\ref{tab::parameters}.

An in-vacuum Andor Ikon CCD detector with 2048x2048 pixels 13.5 $\mu$m x 13.5 $\mu$m in size was used for intensity measurements.
Distances from the focus to detector were varied between experiments (see Table~\ref{tab::parameters}).
FLASH was operated in a single bunch mode with 10 Hz repetition rate.
The detector region of interest and binning in the vertical direction were considered to allow 10 Hz readout frequency.
For the beam attenuation a set of aluminium foil filters and additional silicon nitride films 20 mm x 10 mm in size and varying thicknesses were used in front of the detector.
About $2\cdot 10^4$ intensity profiles were recorded for each monochromator setting.
The vertical direction was the dispersion direction of the monochromator and the intensity profiles were projected on the horizontal axis.
The intensity correlation analysis was performed along the horizontal direction.

Additional information was obtained from the spectrometer detector positioned in the exit slit plane of the monochromator.
Spectrometer detector consisted of a scintillating screen (YAG:Ce 0.2) and an intensified CCD (Andor iStar, DH740), equipped with a lens.
The effective pixel size of the detector in the exit slit plane was 19.4 $\mu$m with the point spread function estimated to be about two pixels (FWHM).
The detector was operated at 10 Hz repetition rate.

\section{Results}

\subsection{Spatial correlation analysis}

\begin{figure}
	\includegraphics[width=\linewidth]{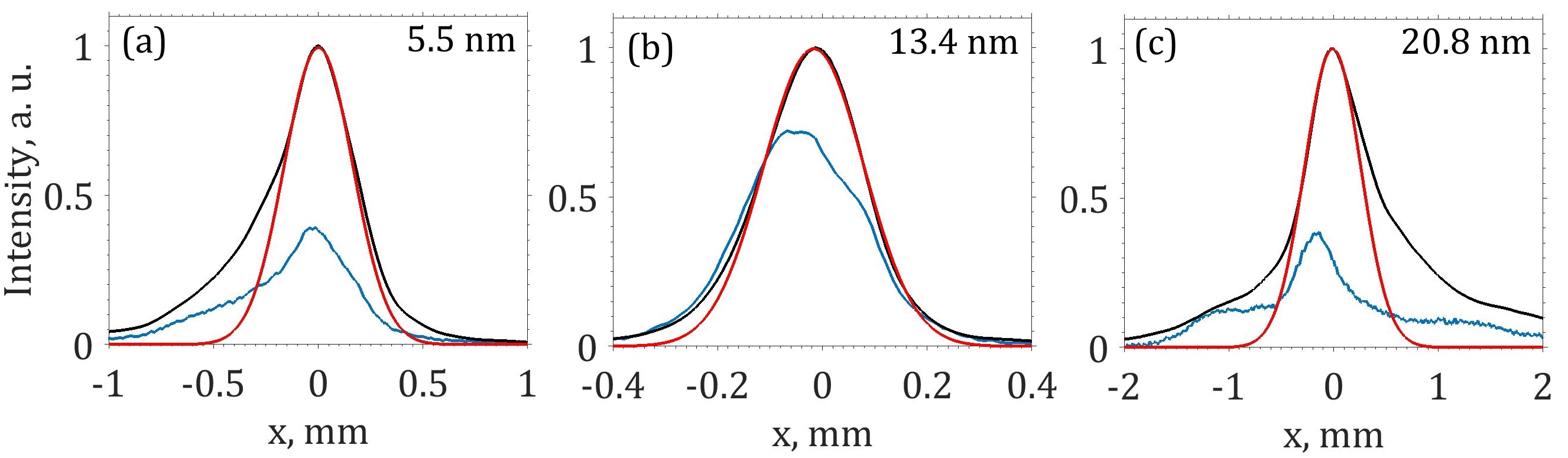}
	\caption{\label{fig::Intens}
		Pulse intensity profiles for 5.5 nm wavelength and $4\cdot 10^{-4}$ bandwidth (a), 13.4 nm wavelength and $1.8\cdot 10^{-4}$ bandwidth (b), 20.8 nm wavelength and $1.8\cdot 10^{-4}$ bandwidth (c).
		Blue lines represent individual pulses, black lines are intensities averaged over $2\cdot10^4$ pulses, red lines are Gaussian fits of a region near the averaged intensity maximum.
	}
	\newpage
\end{figure}

Averaged pulse intensities measured for different FLASH wavelengths and projected on the horizontal direction are presented in Fig.~\ref{fig::Intens}.
The typical individual pulses are also shown in Fig.~\ref{fig::Intens}.
Visual inspection of individual pulses for different modes of operation suggested that in most cases one to two spatial modes were present in FEL beam.
The average pulse intensity can be well represented by a Gaussian fit for $13.4$ nm radiation (see Fig.~\ref{fig::Intens} (b)).
For $5.5$ nm and especially $20.8$ nm wavelength a significant rise of intensity on the beam shoulders can be observed (see Fig. ~\ref{fig::Intens} (a, c)).
The asymmetry of the average intensity suggests inhomogeneity of the FEL radiation properties.
The average horizontal size (FWHM) of the beam was ranging from about $200\ \mu$m to $800\ \mu$m depending on the experiment (see Table~\ref{tab::results})).

\begin{figure}
	\includegraphics[width=\linewidth]{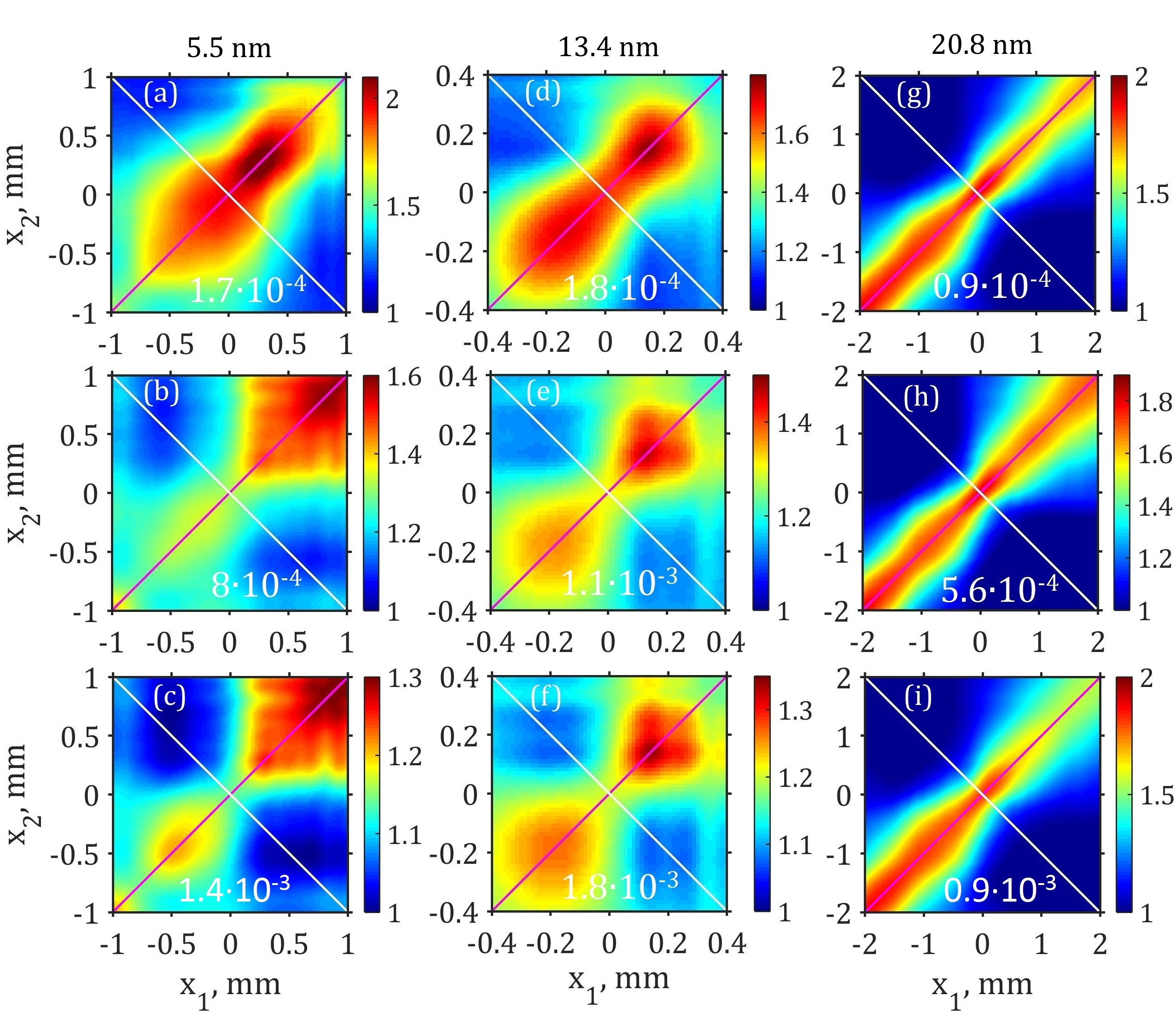}
	\caption{\label{fig::g2}
		Intensity-intensity correlation functions  $g^{(2)}(x_1,x_2)$.
		(a-c) Measurements at 5.5 nm wavelength and the bandwidth $1.7\cdot 10^{-4}$ (a), $8\cdot 10^{-4}$ (b), and $1.4\cdot 10^{-3}$ (c).
        (d-f) Measurements at 13.4 nm wavelength and the bandwidth $1.8\cdot 10^{-4}$ (d), $1.1\cdot 10^{-3}$ (e), and $1.8\cdot 10^{-3}$ (f).
		(g-i) Measurements at 20.8 nm wavelength and the bandwidth $9\cdot 10^{-5}$ (g), $5.6\cdot 10^{-4}$ (h), and $9\cdot 10^{-4}$ (i).
		White lines show the diagonal cross cut over the position of the maximum of intensity $x_1+x_2=0$.
		Purple lines show the autocorrelation direction $x_1 = x_2$.
	}
	\newpage
\end{figure}
\begin{figure}
	\includegraphics[width=\linewidth]{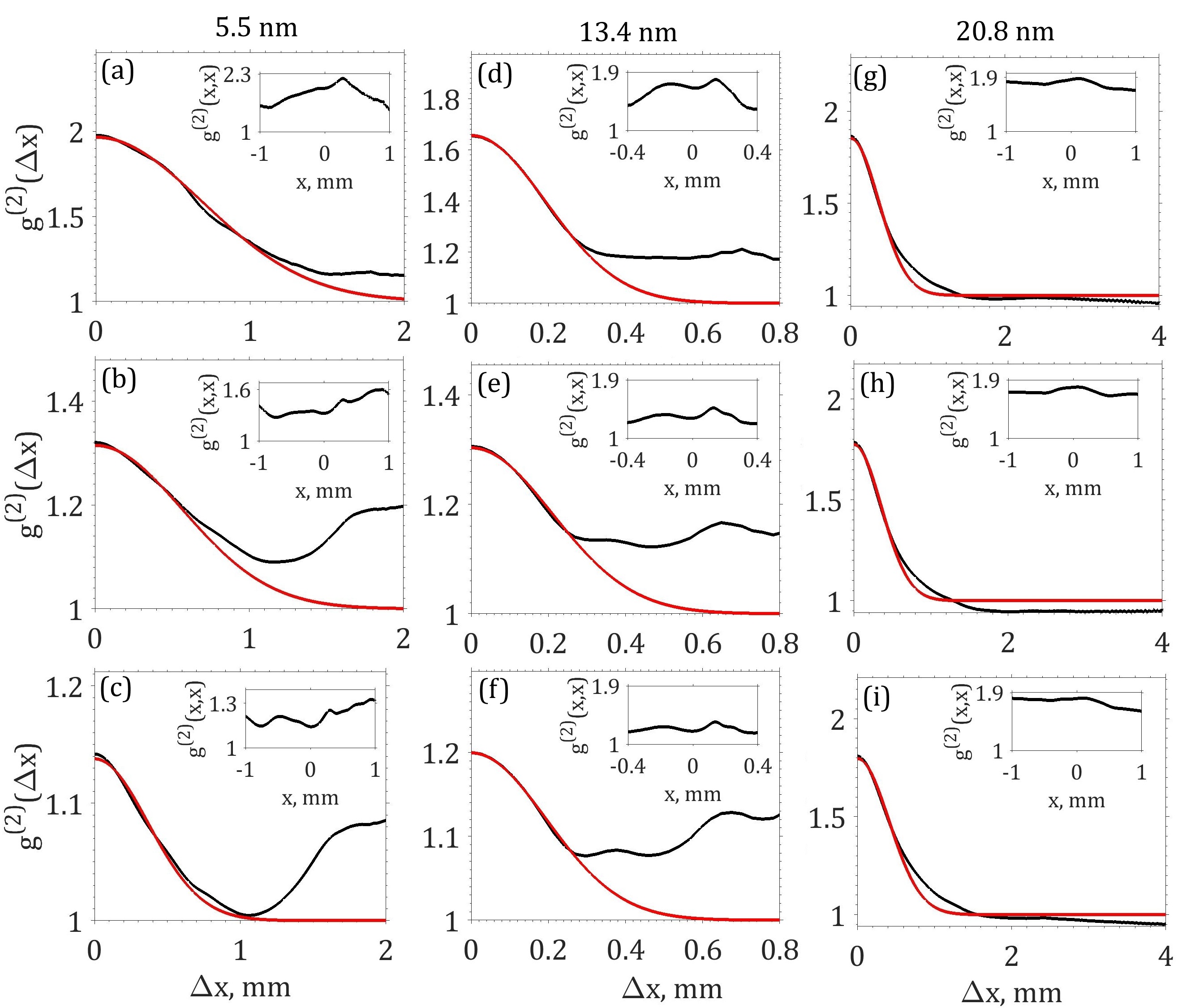}
	\caption{\label{fig::Corr}
		Intensity-intensity correlation function values $g^{(2)}(\Delta x)$ taken along the white lines in Fig.~\ref{fig::g2}  and (insets ) intensity autocorrelation functions
        $g^{(2)}(x,x)$ taken along the purple lines.
		(a-c) Measurements at 5.5 nm wavelength and the bandwidth $1.7\cdot 10^{-4}$ (a), $8\cdot 10^{-4}$ (b) and $1.4\cdot 10^{-3}$ (c).
        (d-f) Measurements at 13.4 nm wavelength and the bandwidth $1.8\cdot 10^{-4}$ (d), $1.1\cdot 10^{-3}$ (e) and $1.8\cdot 10^{-3}$ (f).
        (g-i) Measurements at 20.8 nm wavelength and the bandwidth $9\cdot 10^{-5}$ (g), $5.6\cdot 10^{-4}$ (h) and $9\cdot 10^{-4}$ (i).
		Red lines represent the Gaussian fits.
	}
	\newpage
\end{figure}

Normalized intensity-intensity correlation functions $g^{(2)}(x_1,x_2)$ for all experiments at different bandwidths are shown in Fig.~\ref{fig::g2}.
As is well seen  in this figure, the form of $g^{(2)}(x_1,x_2)$ function is quite complicated and is different from what can be expected for a single Gaussian source (see Fig.~\ref{fig::One_source}).
Intensity correlation function $g^{(2)}(x_1,x_2)$ for $20.8$ nm radiation (see Figs.~\ref{fig::g2} (g-i)) most closely resembles the expected shape for a single chaotic source described by the GSM (compare with Fig.~\ref{fig::One_source}).
At the same time the structure of $g^{(2)}(x_1,x_2)$ at $5.5$ nm (Fig.~\ref{fig::g2} (a-c)) and $13.4$ nm (Fig.~\ref{fig::g2} (d-f)) is more complicated and the presence of several maxima for larger bandwidths along the diagonal $x_2 = x_1$ can be observed (see also insets in Fig.~\ref{fig::Corr}).


To determine the spatial coherence of the FLASH source at different modes of operation we analyzed the diagonal cut of the $g^{(2)}(x_1,x_2)$ function at the position of the intensity maximum shown by white lines in Fig.~\ref{fig::g2} and presented in Fig.~\ref{fig::Corr}.
According to Eq.~\eqref{coherence::g2_traditional} for a single chaotic source these profiles are proportional to square modulus of the spectral degree of coherence $|\mu(x_0-\Delta x/2,x_0+\Delta x/2)|^2$, where $x_0$ is the position of the intensity maximum.
We fitted all profiles with the Gaussian function $1+\zeta_2\exp(-(\Delta x/l_c)^2)$ that provided the values of coherence length $l_c$ for different bandwidths and wavelengths.
We present mean values of the coherence length for all three wavelengths in Table~\ref{tab::results}.
We noticed that the best Gaussian fit for all measured separations $\Delta x$ was obtained for the wavelength of 20.8 nm (see Fig.~\ref{fig::Corr}(g-i)).
The coherence length for this wavelength practically did not depend on the bandwidth value.
Different behavior of the intensity-intensity correlation function was observed at $5.5$ nm (Fig.~\ref{fig::Corr}(a-c)) and $13.4$ nm (Fig.~\ref{fig::Corr}(d-f)) wavelengths at large separations $\Delta x$.
For example, for $5.5$ nm radiation the second maxima appears at larger bandwidths.
This indicates that at these conditions of FLASH operation we have two radiating sources as shown in Fig.~\ref{fig::Different_sources}(b) and Fig.~\ref{fig::Two_angled}.
At the same time for $13.4$ nm we observed a constant background appearing at large separations $\Delta x$.

Comparison of the coherence lengths with the intensity beam size (see Table~\ref{tab::results})) for corresponding experiments showed that the coherence length was about twice larger than the beam size for $5.5$ nm radiation, about the same value for $13.4$ nm and smaller than the beam size for $20.8$ nm wavelength
\footnote{Notice that coherence length is defined as a $\sigma$ value of the distribution and not its FWHM}.
This observation suggests that in the case of $5.5$ nm FEL radiation was noticeably more coherent than in the case of $20.8$ nm.

\begin{figure}
	\includegraphics[width=\linewidth]{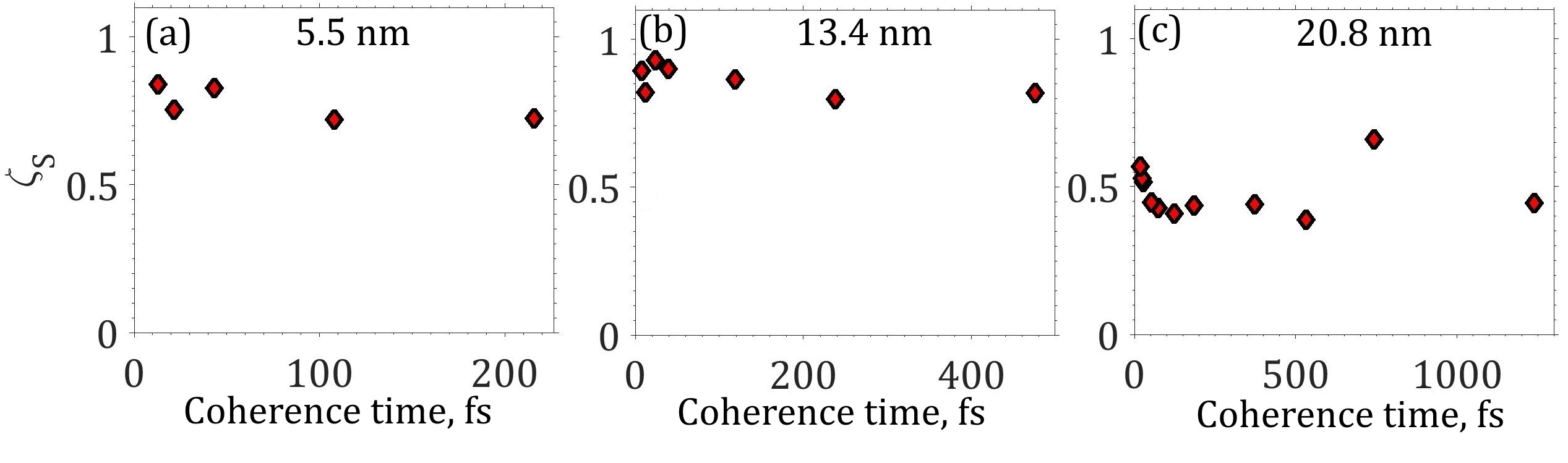}
	\caption{\label{fig::Degree_coherence}
		Degree of spatial coherence as a function of coherence time determined from Eq. \eqref{coherence::coherence_degree_expressed} at 5.5 nm (a), 13.4 nm (b), and 20.8 nm wavelength (c).
	}
	\newpage
\end{figure}

As it follows from our analysis the intensity-intensity correlation function $g^{(2)}(x_1,x_2)$ shows strongly non-uniform behavior (see Fig.~\ref{fig::g2}).
In such situation the global degree of spatial coherence defined in Eq.~\eqref{coherence::coherence_degree_definition} is an adequate measure of spatial coherence.
To determine its values we performed calculations according to Eq. \eqref{coherence::coherence_degree_expressed}.
The degree of spatial coherence as a function of coherence time is presented in Fig.~\ref{fig::Degree_coherence} and mean values are provided in Table~\ref{tab::results}.
It is well seen from this figure that the degree of spatial coherence practically does not depend on the value of coherence time (bandwidth).
It is especially high (above $80\%$) for $5.5$ nm and $13.4$ nm radiation and drops to $50\%$ at $20.8$ nm wavelength.
We want to point out that such high values of the degree of spatial coherence at these radiation wavelengths can be observed only at FELs.
They are about two orders of magnitude higher than for synchrotron sources \cite{Singer2014,Skopintsev2014}.

\begin{figure}
	\includegraphics[width=\linewidth]{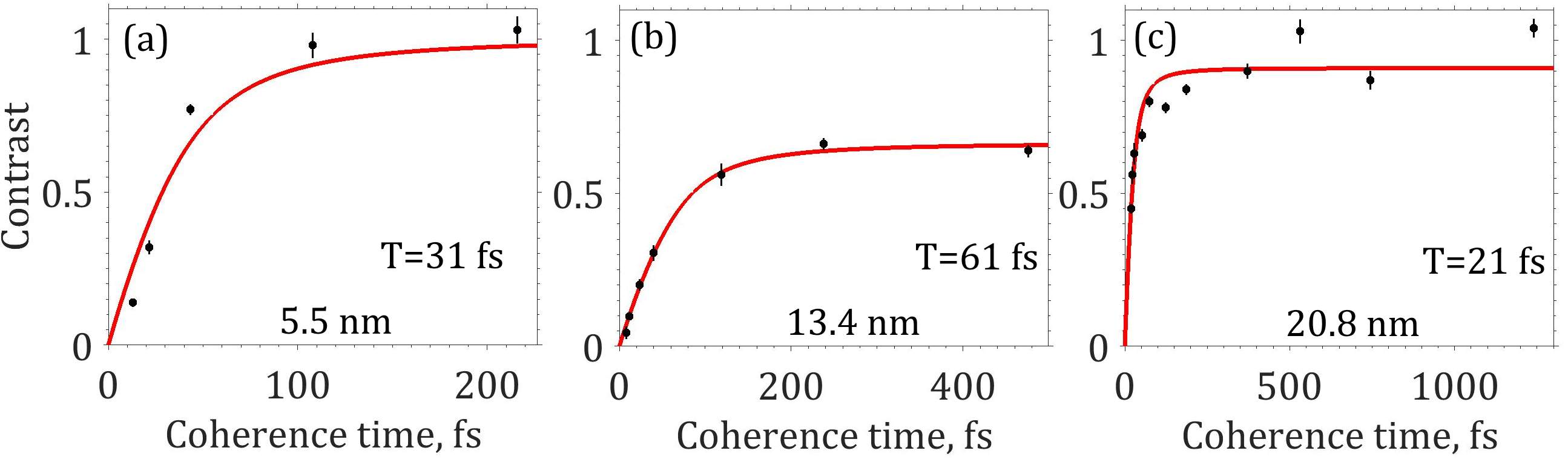}
	\caption{\label{fig::Contrasts}
		Contrast values as a function of coherence time at 5.5 nm (a), 13.4 nm (b), and 20.8 nm wavelength (c).
		Black dots represent experimental points. Errors are calculated as a standard deviation of the contrast in the $1\sigma$ region of intensity. Red line is the fit according to Eq.~\eqref{coherence::contrast_duration_fit}, normalized to the maximum contrast value.
	}
	\newpage
\end{figure}

It is well seen from Fig.~\ref{fig::Corr} that the contrast values (that are maximum values of $g^{(2)}(\Delta x)$ at $\Delta x=0$) strongly depend on the bandwidth
\footnote{Additional variations in contrast may appear due to long term intensity drifts during the measurements.
This can lead to an apparent increase in a contrast (see Appendix for details}.
Variation of contrast $\zeta_2(D_{\omega})$ as a function of coherence time for different wavelengths is shown in Fig.~\ref{fig::Contrasts}.
It displays a typical behavior as described for a chaotic source in Section II.
For large values of coherence time ($\tau_c \gg T$) it reaches a constant value close to unity that indicates that in this case we observe a single longitudinal mode of FEL radiation.
In the opposite limit of small values of coherence time ($\tau_c \ll T$) it showes a linear dependence ($~\tau_c/T$).
Fitting these curves with the theoretical function given in Eq. \eqref{coherence::contrast_duration_fit} provided us with the averaged values of pulse duration ranging from $20$ fs to $60$ fs (see Table~\ref{tab::results}) for the different experiments.
We have to note here that in some cases the contrast values do not reach the maximum value of one at large coherence times (see, for example, Fig.~\ref{fig::Contrasts} (b)).
The reason for that may be a more complicated pulse structure of FEL radiation in deep saturation.
This behavior of the contrast function can be adequately accounted for by normalizing Eq. \eqref{coherence::contrast_duration_fit} by a factor smaller than one.
The values of pulse duration obtained from the analysis of contrast values will be compared in the following with the values obtained from the correlation analysis in the frequency domain.


\subsection{Spectral analysis}

\begin{figure}
	\includegraphics[width=\linewidth]{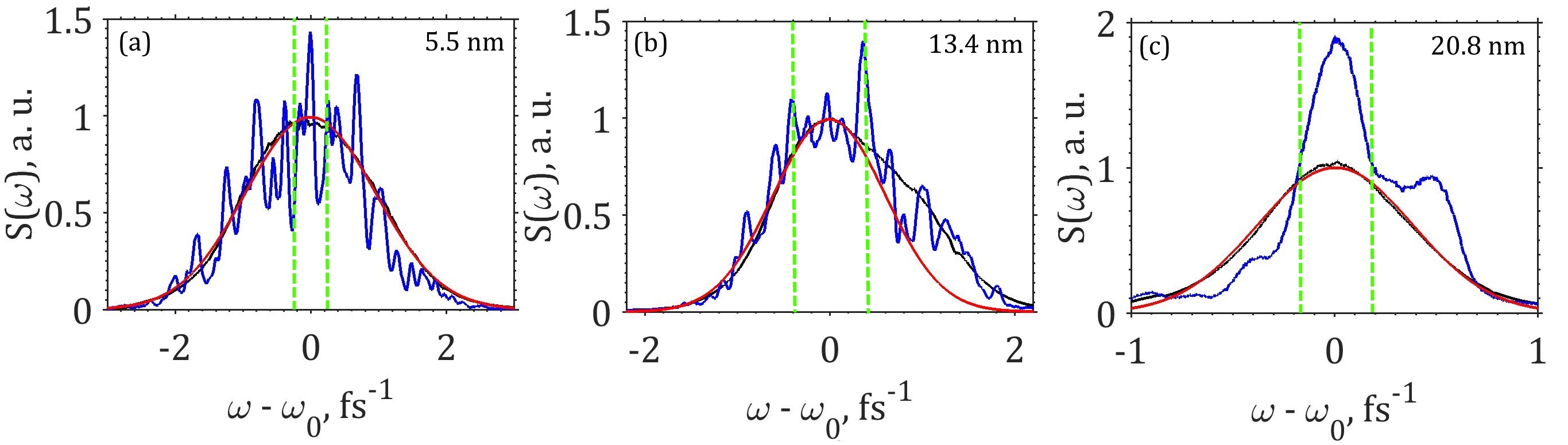}
	\caption{\label{fig::Spectra}
		Spectral measurements performed at PG2 monochromator at 5.5 nm (a), 13.4 nm (b), and 20.8 nm wavelength (c). Blue line is a spectrum of a typical single pulse, black line is the average spectrum, red line is a Gaussian fit to the average spectrum. Green windows depict the bandwidth corresponding to the maximum exit slit opening in each experiment.
	}
	\newpage
\end{figure}

Typical single and average pulse spectra measured in all three experiments are shown in Fig.~\ref{fig::Spectra}.
Longitudinal mode structure is clearly visible in single pulse spectral intensity distributions at all wavelengths.
Note that the largest amount of these modes was observed at $13.4$ nm wavelength, which is predominantly due to longer pulse duration in this experiment.
We can observe also that the modes are overlapping, especially in the case of $13.4$ nm, which is consistent with the value of contrast not reaching unity in Fig.~\ref{fig::Contrasts}.
Our analysis has shown that the average spectrum was very close to Gaussian, especially for $5.5$ nm and $20.8$ nm (see Fig.~\ref{fig::Spectra}).
However, in the experiment at $13.4$ nm wavelength, a deviation of the average spectrum from the Gaussian becomes noticeable at one of the shoulders of the spectrum.



\begin{figure}
	\includegraphics[width=\linewidth]{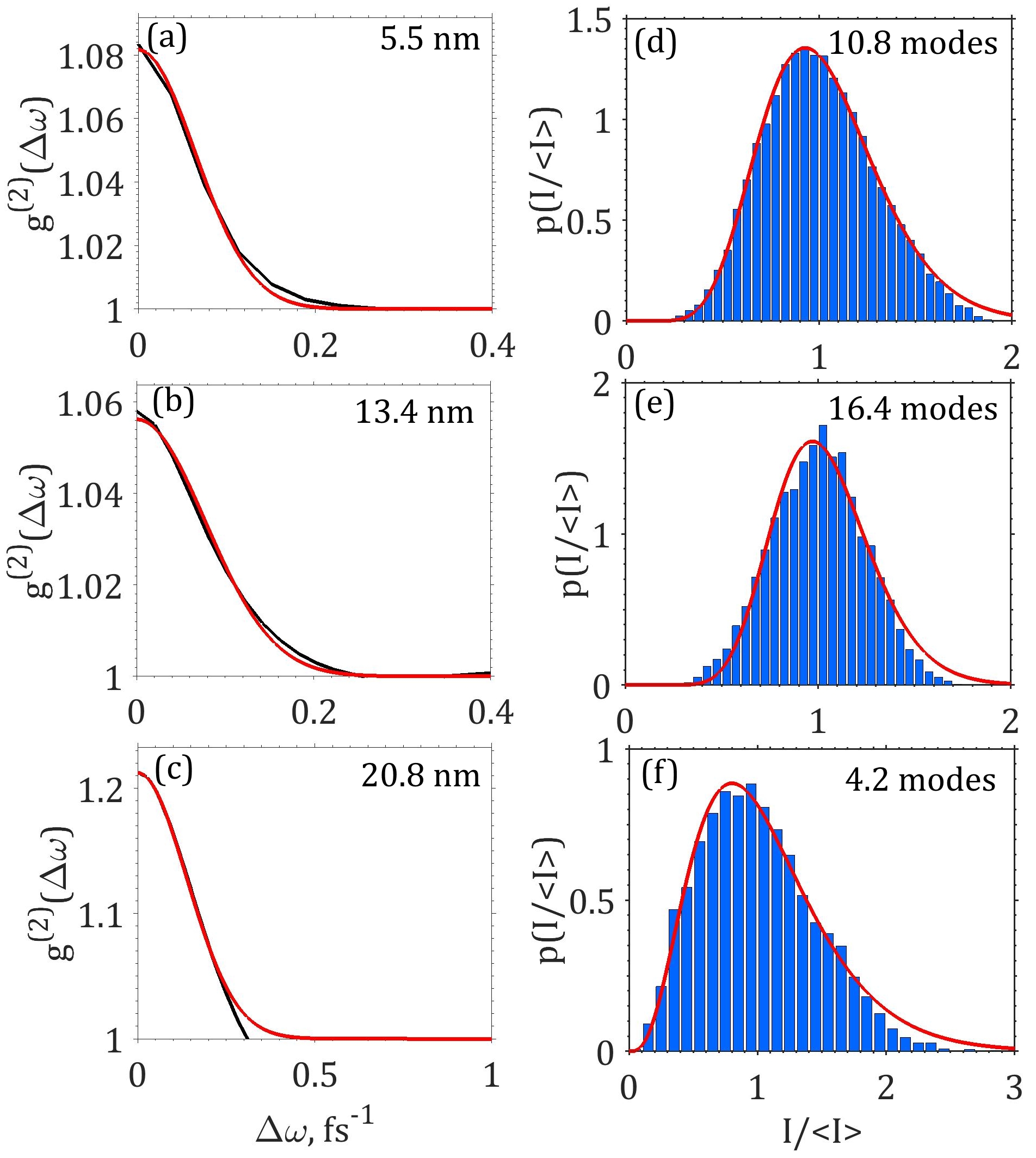}
	\caption{\label{fig::Spectrometer}
		(a-c) Second-order intensity-intensity correlation function $g^{(2)}(\Delta\omega)$ in frequency domain at 5.5 nm (a), 13.4 nm (b), and 20.8 nm wavelength (c).
		Red line is a fit by a Gaussian function.
		(d-f) Histogram of intensity at fixed position on the spectrometer at 5.5 nm (d), 13.4 nm (e), and 20.8 nm (f) wavelength.
		Red line is a fit by a Gamma distribution with a corresponding number of modes.
	}
	\newpage
\end{figure}

It is possible to obtain an estimate of a pulse duration from the analysis of a series of single pulse spectra by the method described in Refs. \cite{Serkez2012, Lutman2012, Engel2016, Singer2013}.
The intensity-intensity correlation function in the spectral domain $g^{(2)}(\omega_1,\omega_2)$ can be defined similar to Eq.~\eqref{coherence::2_order_normalized} as
\begin{equation}
g^{(2)}(\omega_1,\omega_2) = \frac{\langle I(\omega_1 - \omega_0) I(\omega_2 - \omega_0) \rangle}{\langle I(\omega_1 - \omega_0)\rangle \langle I(\omega_2 - \omega_0) \rangle}\ ,
\label{spectra::g2_defin}
\end{equation}
where $\omega_0$ is the central frequency.
Performing such analysis for all three wavelengths and considering on average $10^4$ pulses we determined the second order correlation function $g^{(2)}(\Delta\omega)$ as a function of frequency difference $\Delta\omega = \omega_2 - \omega_1$ (see Fig.~\ref{fig::Spectrometer}(a-c)).
The 2-nd order intensity correlation function determined in this way was fitted by a Gaussian function of the form~\cite{Saldin1998}
\begin{equation}
g^{(2)}(\Delta\omega) = 1+\exp(-\Delta\omega^2\sigma_{T'}^2) \ .
\label{spectra::g2_T}
\end{equation}
This procedure allowed us to estimate an average pulse duration (FWHM) as $T'=2.355 \ \sigma_{T'}$ (see Gaussian fit shown by red lines in Fig.~\ref{fig::Spectrometer}(a-c)).

To determine correct photon pulse duration from our statistical analysis it is also necessary to take into account an effect of the temporal energy chirp of the electron bunch.
Assuming that the energy chirp introduces a linear "chirp-correction" factor \cite{Serkez2012} to the photon pulse duration $T'$ determined by Eq.~\eqref{spectra::g2_T}, we can retrieve a corrected photon pulse duration $T$ as
\begin{equation}
T \approx T'\sigma'_{\omega}/\sigma_{\omega},
\label{spectra::g2_}
\end{equation}
where $\sigma_{\omega}$ is the FEL gain bandwidth and $\sigma'_{\omega}$ is the spectral bandwidth of the FEL radiation.
In our calculations FEL gain bandwidth was considered to be $4\cdot10^{-3} \omega_0$~\cite{Engel2015}.

Pulse durations obtained by spectral analysis described in this section for all wavelengths are given in Table~\ref{tab::results}.
Comparison between the averaged pulse durations obtained by this method and the one based on analysis of the spatial intensity correlation functions described in the previous section shows a good agreement.

Typical histograms of the pulse intensities for different wavelengths are shown in Fig.~\ref{fig::Spectrometer} (d-f).
Pulse intensities were evaluated at a spectral position corresponding to a certain frequency.
The fitting performed with a gamma-function gives best results with $10.8 \pm 0.4$ modes for $5.5$ nm, $16.4 \pm 1.7$ modes for $13.4$ nm and $4.2 \pm 0.3$ modes for $20.8$ nm wavelengths, respectively.
As mentioned in Section II the number of modes is inversely proportional to the contrast value $g^{(2)}(\Delta\omega=0)$.
We observed similar behavior for FLASH radiation with the number of modes obtained from the analysis of the intensity histograms being very close to the inverse contrast values determined by the spatial HBT analysis (see previous sub-section).

\section{Advanced theoretical model}

Comparison between experimental measurements of the intensity-intensity correlation function $g^{(2)}(x_1,x_2)$ observed in Fig.~\ref{fig::g2} and conventional theoretical predictions presented in Fig.~\ref{fig::One_source} and Fig.~\ref{fig::Two_angled} shows significant difference.
To explain this difference an advanced theoretical model will be necessary.
In the following, we introduce two models that consider multiple independent beams and positional beam jitter in the intensity-intensity interferometry.
Situation with the multiple beams at the detector position could appear, for example, if a distant source radiates in different well defined directions (see Fig.~\ref{fig::Different_sources}(c)).
The case of external positional jitter correspond to situation of a distant source radiating randomly in different directions (see Fig.~\ref{fig::Different_sources}(d)).
In this section we will derive theoretical foundation for these two models and in the following section we will compare theoretical predictions with the experimental results.

\subsection{Multiple independent beams}

According to superposition principle for $N$ electric fields with the amplitudes $E_i(\mathbf{r},\omega)$ the total field amplitude is
\begin{equation}
E_{\Sigma}(\mathbf{r},\omega)=
\sum_{i=1}^{N} E_i(\mathbf{r},\omega)\ .
\label{multiple::wave_superposition}
\end{equation}
We assume in the following that all beams are chaotic and obey Gaussian statistics, then the total wave field $E_{\Sigma}(\mathbf{r},\omega)$ also obeys Gaussian statistics \cite{Goodman,MandelWolf}.
As such, we can apply Gaussian moment theorem to the total field and obtain similar to Section II expression for the fourth-order spectral cross-correlation function
$W^{(4)}_{\Sigma}(\mathbf{r_1}, \mathbf{r_2};\omega_1, \omega_2) = S_{\Sigma}(\mathbf{r_1}, \omega_1)S_{\Sigma}(\mathbf{r_2}, \omega_2)+|W^{(2)}_{\Sigma}(\mathbf{r_1}, \omega_1; \mathbf{r_2}, \omega_2)|^2$,
where
$W_{\Sigma}^{(2)}(\mathbf{r_1},\omega_1,\mathbf{r_2},\omega_2) = \langle E_{\Sigma}^*(\mathbf{r_1}, \omega_1) E_{\Sigma}(\mathbf{r_2}, \omega_2 ) \rangle$ is the second-order spectral cross-correlation function of the total beam and $S_{\Sigma}(\mathbf{r},\omega) = W_{\Sigma}^{(2)}(\mathbf{r},\omega,\mathbf{r},\omega)$ is the corresponding spectral density.
Substituting this expression to Eq.~\eqref{coherence::g2_general} we obtain for the intensity-intensity correlation function $g_{\Sigma}^{(2)}(\mathbf{r}_1,\mathbf{r}_2)$ in the case of multiple beams
\begin{equation}
g_{\Sigma}^{(2)}(\mathbf{r}_1,\mathbf{r}_2)
= 1+\frac{\iint\limits_{-\infty}^{\infty} |T(\omega_1)|^2|T(\omega_2)|^2 |W_{\Sigma}^{(2)}(\mathbf{r_1},\omega_1,\mathbf{r_2},\omega_2)|^2\mbox d\omega_1 \mbox d\omega_2}
{\int\limits_{-\infty}^{\infty}|T(\omega)|^2 S_{\Sigma}(\mathbf{r_1},\omega)\mbox d\omega\int\limits_{-\infty}^{\infty}|T(\omega)|^2 S_{\Sigma}(\mathbf{r_2},\omega) \mbox d\omega}\ ,
\label{multiple::g2_multi_entangled}
\end{equation}
If all beams are statistically independent from each other and cross-spectral pure the total spectral cross-correlation function $W^{(2)}_{\Sigma}(\mathbf{r_1},\omega_1,\mathbf{r_2},\omega_2)$ and spectral density $S_{\Sigma}(\mathbf{r},\omega)$ can be expressed as
\begin{eqnarray}
W^{(2)}_{\Sigma}(\mathbf{r_1},\omega_1,\mathbf{r_2},\omega_2)& = & \sum_{i=1}^{N} W^{(2)}_{i}(\mathbf{r_1},\omega_1,\mathbf{r_2},\omega_2) = \sum_{i=1}^{N}J_i(\mathbf{r_1}, \mathbf{r_2})W_i(\omega_1, \omega_2)\ , \\
S_{\Sigma}(\mathbf{r},\omega)& = & \sum_{i=1}^{N} S_{i}(\mathbf{r},\omega) = \sum_{i=1}^{N}I_i(\mathbf{r})S_i(\omega)\ .
\label{multiple::spectral_cross_super}
\end{eqnarray}

Substituting these expressions into Eq. \eqref{multiple::g2_multi_entangled} we obtain for the intensity-intensity correlation function
\begin{widetext}
	\begin{multline}
	g_{\Sigma}^{(2)}(\mathbf{r_1},\mathbf{r_2})
	=1+\\
	 +\frac{\sum_{i,j=1}^{N}J_{i}(\mathbf{r_1},\mathbf{r_2})J^*_{j}(\mathbf{r_1},\mathbf{r_2})\iint\limits_{-\infty}^{\infty}|T(\omega_1)|^2|T(\omega_2)|^2  W_{i}(\omega_1, \omega_2)W^*_{j}(\omega_1, \omega_2)\mbox d\omega_1 \mbox d\omega_2}
	{\sum_{k, l = 1}^{N}I_{k}(\mathbf{r_1})I_{l}(\mathbf{r_2})\int\limits_{-\infty}^{\infty}|T(\omega_1)|^2 S_{k}(\omega_1)\mbox d\omega_1\int\limits_{-\infty}^{\infty}|T(\omega_2)|^2 S_{l}(\omega_2) \mbox d\omega_2}\ .
	\label{multiple::general_g}
	\end{multline}
\end{widetext}

This expression will be used further in simulations.
In general it can not be reduced to a form similar to Eq. \eqref{coherence::g2_traditional} with separated spatial and spectral components.
However, we can now consider two particular cases where such separation can be performed.

It is possible to obtain expressions similar to Eq. \eqref{coherence::g2_traditional} if either spectral or spatial characteristics of each beam are the same.
In the first case $W_{i}(\omega_1,\omega_2)\equiv W(\omega_1,\omega_2)$ and $S_{i}(\omega)\equiv S(\omega)$ and Eq. \eqref{multiple::general_g} reduces to
\begin{equation}
g_{\Sigma}^{(2)}(\mathbf{r_1},\mathbf{r_2})
=1+\zeta_2(D_{\omega})|\mu_{\Sigma}(\mathbf{r_1},\mathbf{r_2})|^2\ ,
\label{multiple::spectra_same_g}
\end{equation}
where $\zeta_2(D_{\omega})$ is defined as in Eq. \eqref{coherence::zeta_2} and
\begin{equation}
|\mu_{\Sigma}(\mathbf{r_1},\mathbf{r_2})|^2 =\frac{\sum_{i, j = 1}^{N}J_{i}(\mathbf{r_1},\mathbf{r_2})J^*_{j}(\mathbf{r_1},\mathbf{r_2})}
{\sum_{i, j = 1}^{N}I_{i}(\mathbf{r_1})I_{j}(\mathbf{r_2})}\ .
\label{multiple::zeta_2}
\end{equation}
In this case it is easy to see that autocorrelation function $g^{(2)}(\mathbf{r},\mathbf{r})$ will be constant, similar to a single chaotic beam (see
Eq.~\eqref{coherence::g2_traditional})
\begin{equation}
g_{\Sigma}^{(2)}(\mathbf{r},\mathbf{r}) = 1+\zeta_2(D_{\omega})\frac{\sum_{i,j=1}^{N}|I_{i}(\mathbf{r})I_{j}(\mathbf{r})|}
{\sum_{i, j = 1}^{N}I_{i}(\mathbf{r})I_{j}(\mathbf{r})} = 1+\zeta_2(D_{\omega})\ .
\label{multiple::spectra_same_auto}
\end{equation}

We assume now that spatial statistical properties of all beams are the same, $J_{i}(\mathbf{r_1},\mathbf{r_2})\equiv J(\mathbf{r_1},\mathbf{r_2})$ and $S_{i}(\mathbf{r}) \equiv S(\mathbf{r})$ but spectral properties are different.
Then again the intensity-intensity correlation function has a form of Eq.~\eqref{coherence::g2_traditional}
\begin{equation}
g^{(2)}(\mathbf{r_1},\mathbf{r_2})
= 1 + \zeta_{2, \Sigma}(D_{\omega})|\mu(\mathbf{r_1},\mathbf{r_2})|^2\ ,
\label{multiple::spatial_same_g}
\end{equation}
where the cross-spectral density function $\mu(\mathbf{r_1},\mathbf{r_2})$ is defined as in Eq. \eqref{coherence::mu2} and the contrast is equal to
\begin{equation}
\zeta_{2, \Sigma}(D_{\omega}) =\frac{\sum_{i,j=1 }^{N}\iint\limits_{-\infty}^{\infty}|T(\omega_1)|^2|T(\omega_2)|^2  W_{i}(\omega_1, \omega_2)W^*_{j}(\omega_1, \omega_2)\mbox d\omega_1 \mbox d\omega_2}
{\sum_{i, j = 1}^{N}\int\limits_{-\infty}^{\infty}|T(\omega_1)|^2 S_{i}(\omega_1)\mbox d\omega_1\int\limits_{-\infty}^{\infty}|T(\omega_2)|^2 S_{j}(\omega_2) \mbox d\omega_2}\ .
\label{multiple::zeta_sum}
\end{equation}
Note, that autocorrelation function $g^{(2)}(\mathbf{r},\mathbf{r})$ in this case is again constant.
Therefore, we can conclude that spatial inhomogeneity in the intensity-intensity correlation function $g^{(2)}(\mathbf{r_1},\mathbf{r_2})$ may appear only if both spatial and spectral properties of different beams are different.

\subsection{Positional jitter}

For a SASE FEL, intensity of each pulse is a random realization of the spatial modes, that leads to the jitter of the center of mass of the intensity distribution from pulse to pulse.
This we will call an {\it intrinsic} jitter.
FEL operation or beamline instabilities may introduce an additional positional jitter to the pulse, which could affect the pulse intensity distribution.
This we will call an {\it external} positional jitter.
It is extremely important and challenging to distinguish between these two effects and, as we will show here, it can be accomplished by the HBT analysis.

We assume here that each pulse at the detector position is randomly shifted by some distance $\mathbf{a}$, which is described by the probability distribution $p(\mathbf{a})$.
Note that by definition this jitter is statistically independent from the statistical distribution of the pulse intensity $I(\mathbf{r})$.
It can be shown (see Appendix for details) that the average intensity distribution due to external jitter $\langle I(\mathbf{r}-\mathbf{a})\rangle$ has a form
\begin{equation}
\langle I_{jit}(\mathbf{r}-\mathbf{a})\rangle = p(\mathbf{r})*\langle I(\mathbf{r})\rangle \ ,
\end{equation}
where $*$ is a convolution operator.
This can be interpreted as a broadening of the intensity due to the external positional jitter.

According to Eq. \eqref{coherence::2_order_normalized} the intensity-intensity correlation function $g^{(2)}_{jit}(\mathbf{r_1}, \mathbf{r_{2}})$ in this case will be modified to the following form
\begin{equation}
g^{(2)}_{jit}(\mathbf{r_1}, \mathbf{r_{2}}) = \frac{\langle I_{jit}(\mathbf{r_1}-\mathbf{a}) I_{jit}(\mathbf{r_2}-\mathbf{a}) \rangle}
{\langle I_{jit}(\mathbf{r_1}-\mathbf{a})\rangle \langle I_{jit}(\mathbf{r_2}-\mathbf{a}) \rangle}\ .
\label{jitter::2_order_normalized}
\end{equation}
It can be further expressed as (see Appendix for details)
\begin{widetext}
	\begin{multline}
g^{(2)}_{jit}(\mathbf{r_1}, \mathbf{r_{2}}) =
\frac{p(\mathbf{r_1})\delta(\mathbf{r_2}-\mathbf{r_1}) * \langle I(\mathbf{r_1})I(\mathbf{r_2})\rangle}
{ \Big[p(\mathbf{r_1})*\langle I(\mathbf{r_1})\rangle\Big] \cdot \Big[p(\mathbf{r_2})*\langle I(\mathbf{r_2})\rangle\Big] }
= \\
= \frac{\Big[ p(\mathbf{r_1})\delta(\mathbf{r_2}-\mathbf{r_1}) \Big] * \Big[ g^{(2)}(\mathbf{r_1}, \mathbf{r_{2}}) \langle I(\mathbf{r_1})\rangle \langle I(\mathbf{r_2})\rangle \Big]}
{ \Big[p(\mathbf{r_1})*\langle I(\mathbf{r_1})\rangle\Big] \cdot \Big[p(\mathbf{r_2})*\langle I(\mathbf{r_2})\rangle\Big] }  \ ,
\label{jitter::2_order_normalized_expr}
	\end{multline}
\end{widetext}
where $\delta(x)$ is the $\delta$-function.
By this we expressed intensity-intensity correlation function due to external jitter $g^{(2)}_{jit}(\mathbf{r_1}, \mathbf{r_{2}})$ through the correlation function $g^{(2)}(\mathbf{r_1}, \mathbf{r_{2}})$ without jitter.

The autocorrelation function can be derived from Eq.~\eqref{jitter::2_order_normalized_expr} as
\begin{equation}
g^{(2)}_{jit}(\mathbf{r}, \mathbf{r}) = \frac{p(\mathbf{r})*\langle I^2(\mathbf{r})\rangle}{ \Big[p(\mathbf{r})*\langle I(\mathbf{r})\rangle\Big]^2}\ .
\label{jitter::autocorr}
\end{equation}

As it follows from these results an external jitter could significantly change statistics of the measured pulses.
If original intensity distribution obeys Gaussian statistics, as it follows from Eqs.~\eqref{jitter::2_order_normalized_expr} and \eqref{jitter::autocorr} it is not necessarily the case if an external jitter is present.
This means also that expression \eqref{coherence::g2_traditional} is not any more valid in this situation.

Obtained Eqs.~\eqref{jitter::2_order_normalized_expr} and \eqref{jitter::autocorr} are quite general, below we will estimate an effect of jitter on the intensity-intensity autocorrelation function in the one-dimensional case (see Appendix for details).
We will assume that the intrinsic autocorrelation function without a jitter is a constant $g^{(2)}_0(x,x)=g^{(2)}_0$, as it is the case for a single chaotic source (see Fig.~\ref{fig::One_source}).
We will consider also that both the average intensity profile of the radiation $\langle I(x)\rangle$ and positional jitter $p(a_x)$ are described by a normal distribution with the corresponding rms values $\sigma_I$ and $\sigma_{jit}$.
Under these conditions we can determine the intensity-intensity autocorrelation function with the jitter as (see Appendix for details)
\begin{equation}
g^{(2)}_{jit}(x, x) = g^{(2)}_0 \cdot \frac{1+k^2}{ \sqrt{1+2k^2}}\exp\left[\frac{x^2}{2\Sigma^2}\right]\ ,
\label{jitter::autocorr_gauss}
\end{equation}
where $\Sigma^2 = \sigma_I^2(1+k^2)(1+2k^2)/(2k^2)$ and $k=\sigma_{jit}/\sigma_I$.
This expression shows that the jitter has a minimum effect on the contrast near the central intensity position, however, contrast grows exponentially at larger deviations from the center of the beam in the case of external positional jitter.

\section{Discussion}

In this section we will apply models of multiple beams and positional jitter developed in the previous section to explain most of the observed features of the intensity-intensity correlation function $g^{(2)}(x_1,x_2)$.

\subsection{Multiple beams}

Two different examples of spatial inhomogeneities in the intensity-intensity correlation function $g^{(2)}(x_1,x_2)$ are well seen in Fig.~\ref{fig::g2}(b-c) and Fig.~\ref{fig::g2}(g-i).
In the first case, large values of the correlation function are observed in the quadrant $x_1>0 \ , x_2>0$. In the second, $g^{(2)}(x_1,x_2)$ is more homogeneous along the direction $x_2=x_1$ however its values are slightly varying.

To simulate such behaviour of intensity correlation function we applied multiple beam model developed in the previous section.
We considered two GSM beams described by Eqs. (\ref{coherence::GSM}-\ref{coherence::GSM_JW}) with the parameters listed in Table~\ref{tab::sim_param} and called "Model II" and "Model III".
Model II is characterized by comparably large separation between two beams, as well as strong difference in the beam relative intensity and values of the coherence length.
In Model III, both beams are also separated for about the same distance but have similar values of the coherence length.
In both models spectral coherence width $\Omega_c$ of both beams is different.

\begin{figure}
	\includegraphics[height=17cm]{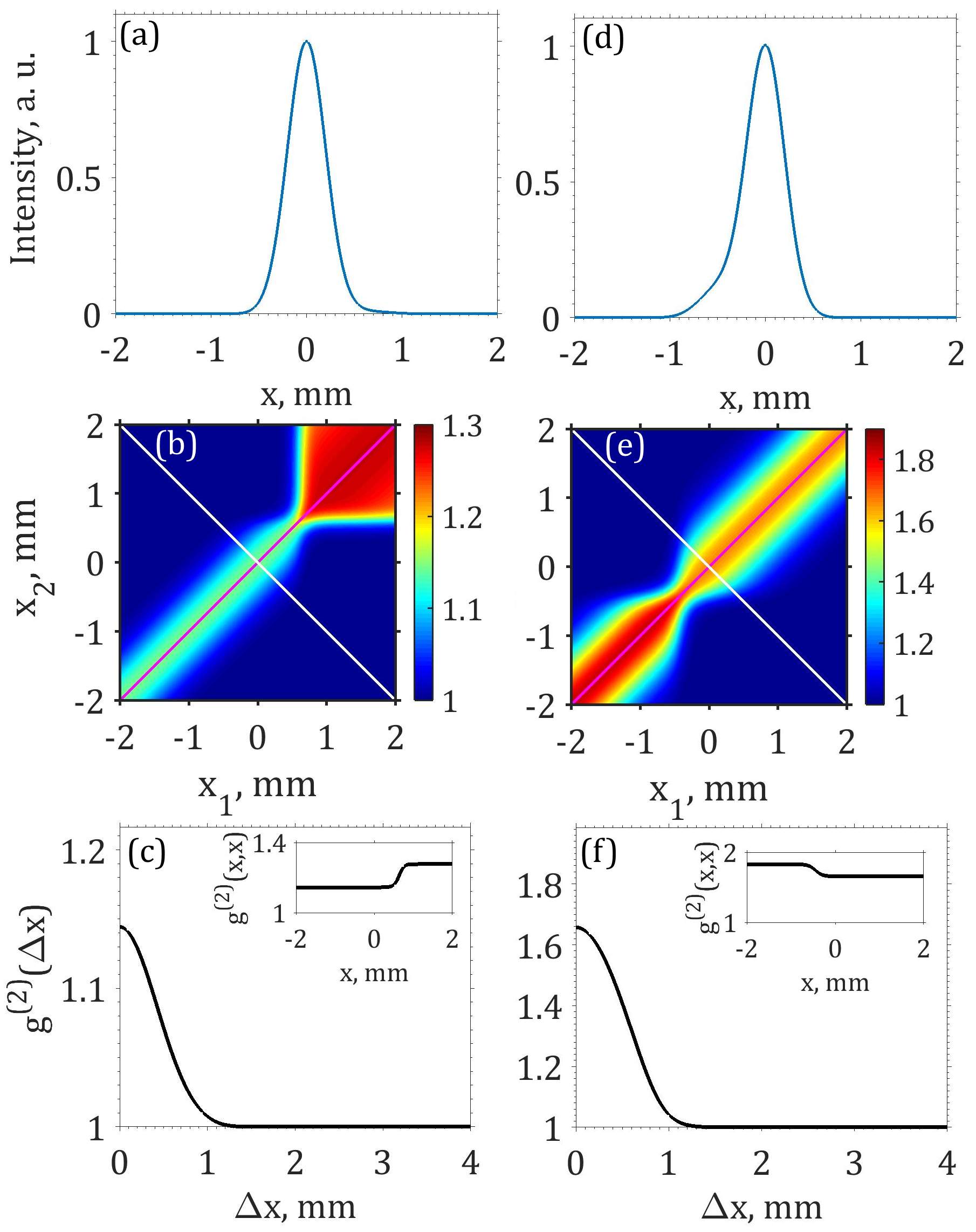}
	\caption{\label{fig::inhomogeneous}
		Results of simulations for two cases of independent beams shown in Table~\ref{tab::sim_param} as Model II (a - c) and Model III (d - f).
        (a, d) Intensity, (b, e) intensity-intensity correlation function $g^{(2)}(x_1,x_2)$, (c, f) values of $g^{(2)}(\Delta x)$ taken along white lines in (b, e). In insets in (c, f) the autocorrelation function $g^{(2)}(x,x)$ taken along purple lines in (b, e) is shown.
	}
	\newpage
\end{figure}

Simulations were performed using general Eq.~\eqref{multiple::general_g}.
Results of these simulations are shown in Fig.~\ref{fig::inhomogeneous}.
As it is well seen from this Figure results of our simulations well reproduce our experimental results shown in Fig.~\ref{fig::g2}.
Model II (Fig.~\ref{fig::inhomogeneous}(a-c)) gives similar results as obtained in Fig.~\ref{fig::g2} (b-c) and Model III (Fig.~\ref{fig::inhomogeneous}(d-f)) reproduces experimental results shown in Fig.~\ref{fig::g2}(g-i).
For example, we see that the autocorrelation function (shown in the insets of Figs.~\ref{fig::inhomogeneous}(b,d)) is inhomogeneous.
We also observe a certain narrowing of the intensity correlation function between two beams that effectively decreases the coherence length in this region (see Figs.~\ref{fig::inhomogeneous}(b,e)).
Far away from the region of the beams overlap characteristics of the beams coincide with their original values.
As such, we can explain inhomogeneities observed in our analysis by the presence of multiple beams measured on the detector.
Here we do not discuss the origin of these beams.
It can be due to lasing of different parts of the electron bunch, or due to rescattering by the optical elements present in the beamline.
More detailed investigation of these effects may become a separate project in future.

\subsection{Positional jitter}

\begin{figure}
	\includegraphics[height=17cm]{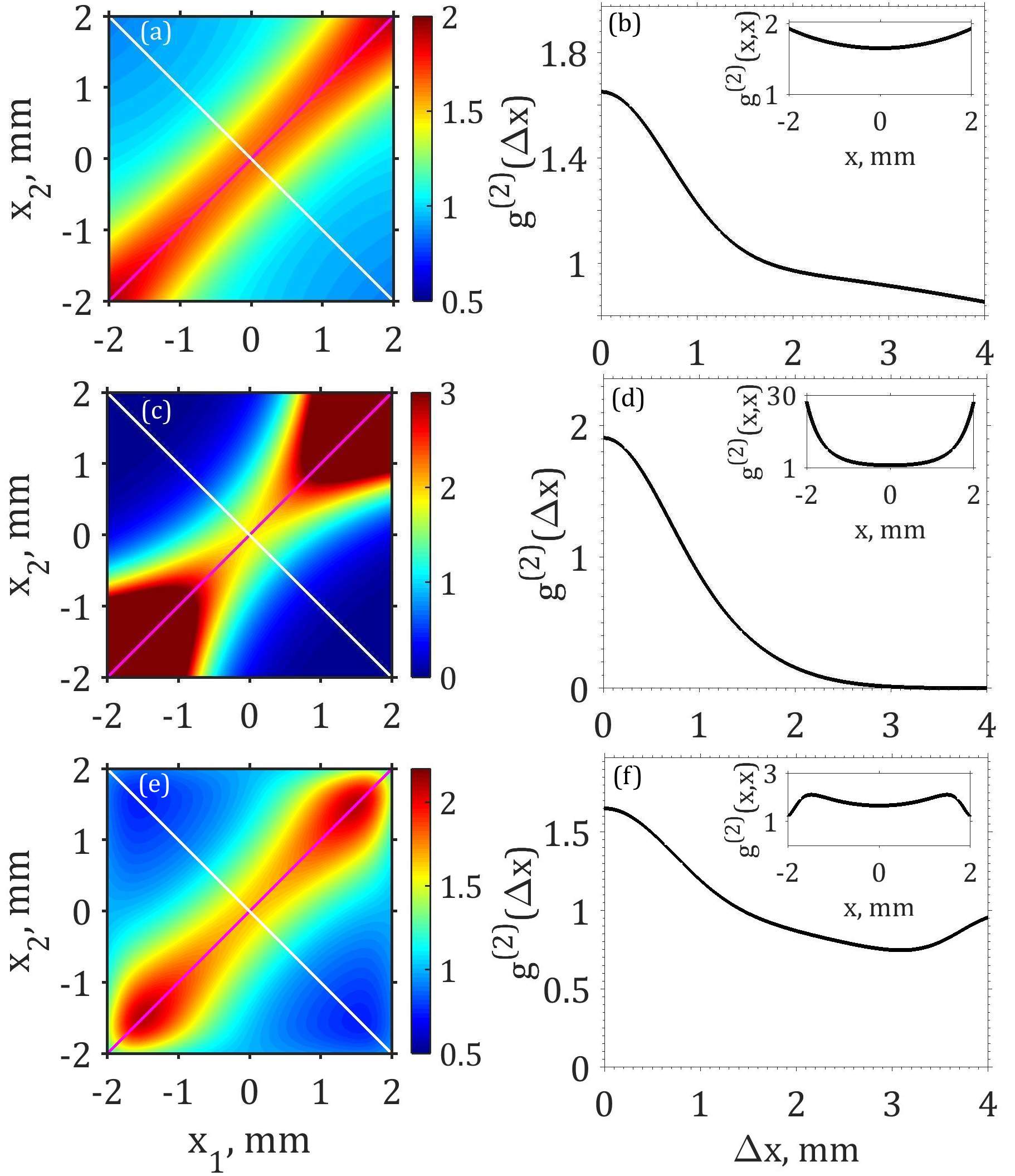}
	\caption{\label{fig::jitter}
		(a,c) Intensity-intensity correlation function $g^{(2)}(x_1,x_2)$ with different values of the positional jitter calculated according to Eq.~\eqref{jitter::2_order_normalized_expr}.
        Ratio of the positional jitter to the size of the beam was $k=0.1$ (a) and $k = 0.2$ (c).
		(b,d) Intensity-intensity correlation function $g^{(2)}(\Delta x)$ taken along the white lines in Fig.~\ref{fig::jitter}(a,c).
        In the insets the corresponding autocorrelation functions $g^{(2)}(x,x)$ taken along the purple lines are shown.
		(e,f) Same simulations as in (c,d) with additional random noise in the intensity distribution with the dispersion value of 0.1\%.
		Statistical properties of the beam were the same as listed in Table~\ref{tab::sim_param} as Model I.
	}
	\newpage
\end{figure}

Another feature clearly visible in our measurements at $13.4$ nm wavelength (see Fig.~\ref{fig::g2}(e-f)) is the presence of two symmetrical maxima of about the same magnitude along the diagonal $x_2=x_1$.
Such behavior of $g^{(2)}(x_1,x_2)$ function can be explained by the positional jitter of the beam.
We demonstrate this by performing simulations with different levels of the positional jitter according to Eq.~\eqref{jitter::2_order_normalized_expr} as shown in Fig.~\ref{fig::jitter}.
Beam parameters were considered to be the same as described in Table~\ref{tab::sim_param} as Model I.
The positional jitter was described by the normal probability distribution with the value of the parameter $k$: $k=0.1$ in Fig.~\ref{fig::jitter}(a,b) and $k=0.2$ in Fig.~\ref{fig::jitter}(c,d).
In these figures we observe an exponential growth of the autocorrelation function $g^{(2)}(x,x)$ along the main diagonal $x_2=x_1$ in accordance with Eq.~\eqref{jitter::autocorr_gauss} (see insets in Fig.~\ref{fig::jitter}(b,d)).
At the same time the values of the 2-nd order intensity correlation function $g^{(2)}(\Delta x)$ become smaller than unity in the direction along the diagonal $x_2=-x_1$ (see Fig.~\ref{fig::jitter}(b,d)).

In our experiments we did not observe such strong exponential growth of $g^{(2)}(x,x)$ function but rather the presence of two maxima (see Fig.~\ref{fig::g2}(e-f)).
To simulate our experimental results we added random noise with the dispersion value of $0.1\%$ of the maximum intensity to the intensity distribution with the jitter value $k=0.2$ (see Appendix for details).
Results of these simulations are presented in Fig.~\ref{fig::jitter}(e).
Comparison of our experimental results in Fig.~\ref{fig::g2}(e-f) and results of these simulations show that we can reproduce the main features observed in the experiment.
By fitting the autocorrelation function $g^{(2)}(x, x)$ near the center of the beam in Fig.~\ref{fig::g2}(e-f) with Eq.~\eqref{jitter::autocorr_gauss} we obtained an estimate of the  positional jitter in our experiment to be about $25\%$ of the beam size (parameter $k\approx 0.25$).

\section{Summary}

In summary, we performed HBT interferometry measurements at FLASH.
We determined the 2-nd order intensity correlation function at the wavelengths $5.5$ nm, $13.4$ nm, and $20.8$ nm and different operation conditions of the FEL.
In all measurements we obtained high degree of spatial coherence  that was above 50\%.
This is about two orders of magnitude higher then at any synchrotron source in the same energy range.
From our statistical measurements we also extracted an average pulse duration that was below 60 fs.
These results were compared with the measurements of the 2-nd order correlation function in spectral domain that provided independent measurements of the pulse durations.
Both approaches showed close agreement between two methods.

Our results also revealed that the 2-nd order intensity correlation function of FLASH radiation is strongly inhomogeneous and its behavior could not be explained by radiation coming from a single distant extended source.
We developed advanced theoretical models that included multiple beam model and external positional jitter to account for these effects.
By performing simulations we could reproduce the behavior of the 2-nd order intensity correlation function observed in the experiment.
The developed approach allowed us to estimate the magnitude of the external positional jitter that was about $25\%$ of the beam size in the experiment at $13.4$ nm wavelength.

Obtained results demonstrate that the HBT interferometry is a very sensitive method for the FEL beam statistical characterization.
Simple experimental setup allows to perform measurements of the 2-nd order intensity correlation function in parallel with the main experiment, that will be especially useful for the experiments performed in the gas phase, or diluted targets.
Such on-line measurements will allow to monitor FEL operation, provide an important feedback for the machine operation and could potentially become a sensitive diagnostic tool of the FEL performance.

What is even more thrilling that our measurements demonstrated high degree of spatial coherence of the FEL radiation that could potentially lead to completely new avenue in the field of quantum optics.
It opens the route for such quantum optics experiments as exploration of non-classical states of light \cite{Breitenbach1997},
super-resolution experiments \cite{Oppel2012}, or ghost imaging experiments \cite{Pelliccia2016,Yu2016} at the FEL sources.
As it was shown in our work by sufficient monochromatization single longitudinal mode of the FEL radiation, or Fourier limited pulses with high photon flux can be acheaved.
This in turn is an important prerequisite for a new class of light phase sensitive experiments such as coherent and phase control interferometry with attosecond precision \cite{Prince2016, Usenko2016}.
Finally, we foresee that HBT interferometry will become in the near future an important diagnostics and analysis tool at the FEL sources.

\section{Acknowledgments}

We acknowledge the support of the groups of J. von Zanthier and R. R{\"o}hlsberger for the measurements of the 2-nd order intensity correlation function at 13.4 nm wavelength of FLASH during a joint experiment.
We thank E. Weckert for helpful discussions and support of the project.
We thank S. Serkez and E. Saldin for fruitful discussions concerning FEL physics.
We are grateful to the FLASH scientific and technical staff for making the experiment possible.
This work was partially supported by the Virtual Institute VH-VI-403 of the Helmholtz Association.

\clearpage

\appendix

\section{Intensity drifts during the measurements}

\begin{figure}
	\includegraphics[width=\linewidth]{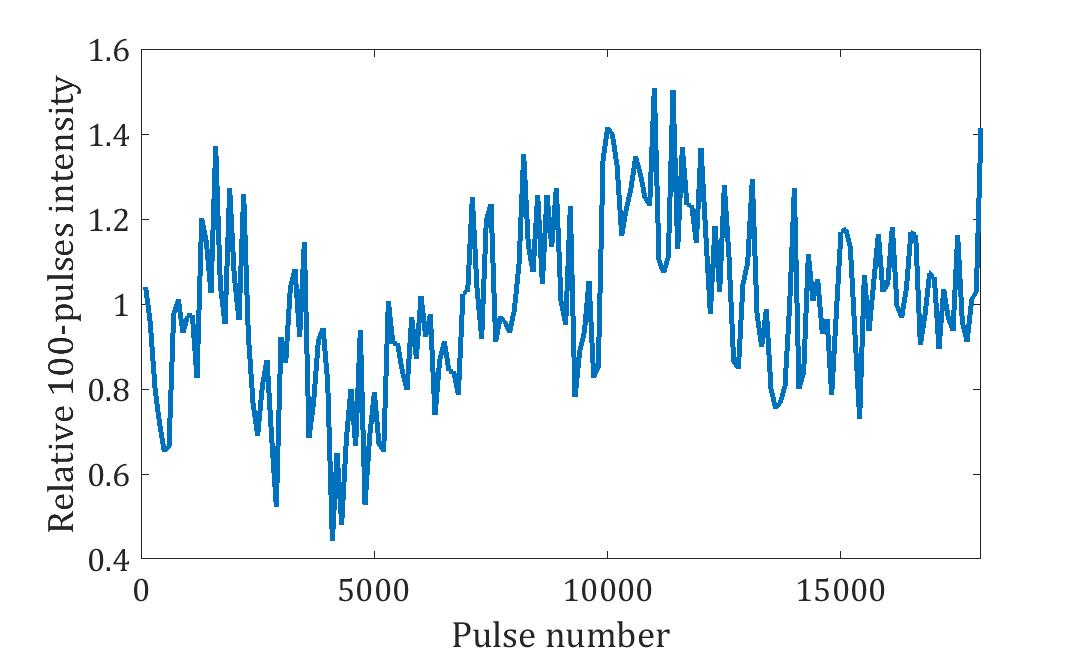}
	\caption{\label{fig::intensity_drift}
		Drift of the relative intensity averaged over 100 pulses during the run.
	}
	\newpage
\end{figure}

Long term intensity drifts during the experiment could also effect the value of the measured contrast.
FLASH intensity drifts during one of the experiments are shown in Figure \ref{fig::intensity_drift}.
Such drifts introduce additional error in the 2-nd order intensity correlation function, since our assumption that each pulse is a different realization of the SASE statistical distribution is violated in this case.



\section{Positional jitter}

HBT interferometry at FELs involves averaging at least over several thousand pulses.
For a chaotic source, intensity center of mass changes its position from pulse to pulse.
In addition to that, external vibrations can introduce external positional jitter to the pulse, which can only be distinguished from the natural jitter of the chaotic radiation by statistical analysis.
In the following subsection we will describe the effect the external jitter has on intensity-intensity correlation function $g^{(2)}(\mathbf{r_1},\mathbf{r_2})$.

Let us assume that every pulse is randomly shifted by some distance $\mathbf{a}$, which is described by probability distribution $p(\mathbf{a})$.
Notice that by definition this jitter is statistically independent from the statistical distribution of pulse intensity $I(\mathbf{r})$.
Intensity-intensity correlation function $g^{(2)}(\mathbf{r_1}, \mathbf{r_{2}})$ will then be, following from Eq. (5) in the main text

\begin{equation}
g^{(2)}_{jit}(\mathbf{r_1}, \mathbf{r_{2}}) = \frac{\langle I_{jit}(\mathbf{r_1}-\mathbf{a}) I_{jit}(\mathbf{r_2}-\mathbf{a}) \rangle}{\langle I_{jit}(\mathbf{r_1}-\mathbf{a})\rangle \langle I_{jit}(\mathbf{r_2}-\mathbf{a}) \rangle}\ .
\label{A_jitter::2_order_normalized}
\end{equation}

Let us first calculate the average intensity $\langle I_{jit}(\mathbf{r}-\mathbf{a})\rangle$.
We can express

\begin{equation}
\langle I_{jit}(\mathbf{r}-\mathbf{a})\rangle = \langle F^{-1}[F(I_{jit}(\mathbf{r}-\mathbf{a}))]\rangle = \left\langle F^{-1}\left[e^{-i\mathbf{q}\mathbf{a}}F(I(\mathbf{r}))\right]\right\rangle \ ,
\end{equation}
where $F$ is the Fourier transform.
We assume here that the Fourier transform is of the form $F(f(\mathbf{r}))=1/(2\pi)^n\int e^{-i\mathbf{q}\mathbf{r}}f(\mathbf{r})d\mathbf{r}$.
Since jitter is statistically independent from $I(\mathbf{r})$, we derive

\begin{equation}
\langle I_{jit}(\mathbf{r}-\mathbf{a})\rangle = F^{-1}\left[\left\langle e^{-i\mathbf{q}\mathbf{a}}\right\rangle F\Big(\langle I(\mathbf{r})\rangle\Big)\right]\ .
\end{equation}

We note here that by definition of Fourier transform $\left\langle e^{i\mathbf{q}\mathbf{a}}\right\rangle = (2\pi)^{n/2} F^{-1}(p(\mathbf{a}))(\mathbf{q})$.
Using the convolution theorem for Fourier transform \cite{FourierBook} we obtain

\begin{equation}
\langle I_{jit}(\mathbf{r}-\mathbf{a})\rangle = \frac{1}{(2\pi)^{n/2}}F^{-1}\left[(2\pi)^{n/2} F^{-1}\left(p(\mathbf{-a})\right)\right]*F^{-1}\left[F\Big(\langle I(\mathbf{r})\rangle\Big)\right] = p(\mathbf{r})*\langle I(\mathbf{r})\rangle \ .
\end{equation}

Here $*$ denotes convolution.
As expected, the average intensity is "smeared" by the positional jitter.
Performing the same operations on $\langle I_{jit}(\mathbf{r_1}-\mathbf{a})I_{jit}(\mathbf{r_2}-\mathbf{a})\rangle$ we obtain

\begin{multline}
\langle I_{jit}(\mathbf{r_1}-\mathbf{a})I_{jit}(\mathbf{r_2}-\mathbf{a})\rangle = \left\langle F^{-1}\left[e^{-i(\mathbf{q_1}+\mathbf{q_2})\mathbf{a}}F\Big(I(\mathbf{r_1})I(\mathbf{r_2})\Big)\right]\right\rangle = \\ = F^{-1}\left[\left\langle e^{-i(\mathbf{q_1}+\mathbf{q_2})\mathbf{a}}\right\rangle F\Big(\langle I(\mathbf{r_1})I(\mathbf{r_2})\rangle\Big)\right]\ ,
\end{multline}
and using convolution theorem again

\begin{multline}
\langle I_{jit}(\mathbf{r_1}-\mathbf{a})I_{jit}(\mathbf{r_2}-\mathbf{a})\rangle = \\ =  \frac{1}{(2\pi)^n}F^{-1}\left[\sqrt{2\pi}^n F_{\mathbf{q_1}+\mathbf{q_2}}^{-1}(p(-\mathbf{a}))\right]*F^{-1}\left[F\Big(\langle I(\mathbf{r_1})I(\mathbf{r_2})\rangle\Big)\right] = \\  = p(\mathbf{r_1})\delta(\mathbf{r_2}-\mathbf{r_1})*\langle I(\mathbf{r_1})I(\mathbf{r_2})\rangle\ .
\end{multline}

Now we can write the expression for intensity-intensity correlation function $g^{(2)}_{jit}(\mathbf{r_1}, \mathbf{r_{2}})$
\begin{equation}
g^{(2)}_{jit}(\mathbf{r_1}, \mathbf{r_{2}}) = \frac{p(\mathbf{r_1})\delta(\mathbf{r_2}-\mathbf{r_1})*\langle I(\mathbf{r_1})I(\mathbf{r_2})\rangle}{ \Big(p(\mathbf{r_1})*\langle I(\mathbf{r_1})\rangle\Big) \cdot \Big(p(\mathbf{r_2})*\langle I(\mathbf{r_2})\rangle\Big) }\ .
\label{A_jitter::2_order_normalized_expr}
\end{equation}

Equation \eqref{A_jitter::2_order_normalized_expr} could be also presented in an integral form as
\begin{equation}
g^{(2)}_{jit}(\mathbf{r_1}, \mathbf{r_{2}}) = \frac{\int p(\mathbf{r})\langle I(\mathbf{r_1-r})I(\mathbf{r_2-r})\rangle \mbox d\mathbf{r}}{ \Big(\int p(\mathbf{r})\langle I(\mathbf{r_1-\mathbf{r}})\rangle \mbox d\mathbf{r}\Big) \cdot \Big(\int p(\mathbf{r})\langle I(\mathbf{r_2-\mathbf{r}})\rangle \mbox d\mathbf{r}\Big) }\ .
\label{jitter::2_order_normalized_integral}
\end{equation}

The autocorrelation function can be derived from Eq.~\eqref{A_jitter::2_order_normalized_expr} as

\begin{equation}
g^{(2)}_{jit}(\mathbf{r}, \mathbf{r}) = \frac{p(\mathbf{r})*\langle I^2(\mathbf{r})\rangle}{ \Big[p(\mathbf{r})*\langle I(\mathbf{r})\rangle\Big]^2}\ .
\label{A_jitter::autocorr}
\end{equation}

As an example we can estimate how the intensity-intensity autocorrelation function will look like in one-dimensional case with jitter.
Let us assume Gaussian intensity profile $\langle I(x)\rangle$ and Gaussian jitter probability distribution $p(a_x)$
\begin{eqnarray}
\label{A_jitter::avi}
\langle I(x)\rangle & = & I_0 e^{-x^2/2\sigma_I^2}\ , \\
\label{A_jitter::avp}
p(a_x) & = & \frac{1}{\sqrt{2\pi}\sigma_{jit}} e^{-a_x^2/2\sigma_{jit}^2}\ .
\end{eqnarray}
We will also assume that intrinsic autocorrelation function is a constant $g^{(2)}(x,x)=g^{(2)}_0$, as it is the case for a single chaotic source.
Then $\langle I^2(x)\rangle = g^{(2)}_0 \langle I(x) \rangle^2$ by definition of $g^{(2)}(x,x)$.
Substituting this relation as well as Eqs.~\eqref{A_jitter::avi}, and \eqref{A_jitter::avp} into Eq.~\eqref{A_jitter::autocorr}, we immediately obtain
\begin{equation}
g^{(2)}_{jit}(x, x) =\sqrt{2\pi}\sigma_{jit} g^{(2)}_0 \frac{ e^{-x^2/2\sigma_{jit}^2}*e^{-x^2/\sigma_I^2}}{ \Big[e^{-x^2/2\sigma_{jit}^2}*e^{-x^2/2\sigma_I^2}\Big]^2}\ .
\label{A_jitter::autocorr_oned}
\end{equation}

Using well-known fact that the convolution of two normalized Gaussian functions with the rms values $\sigma_1$ and $\sigma_2$ is a normalized Gaussian function with the rms value $\sigma = \sqrt{\sigma_1^2+\sigma_2^2}$ (see, for example, \cite{convolution}), we finally obtain for the intensity-intensity autocorrelation function with the jitter
%
%
%
\begin{equation}
g^{(2)}_{jit}(x, x) = g^{(2)}_0 \cdot \frac{1+k^2}{ \sqrt{1+2k^2}}\exp\left[\frac{x^2}{2\Sigma^2}\right]\ ,
\label{A_jitter::autocorr_gauss}
\end{equation}
where $\Sigma^2 = \sigma_I^2(1+k^2)(1+2k^2)/(2k^2)$ and $k=\sigma_{jit}/\sigma_I$.

\section{Effects of the background noise}

The intensity-intensity correlation function $g^{(2)}(x_1, x_2)$ in the region of low intensity values can be affected by the background noise.
We assume that the total intensity can be represented as
\begin{equation}
I(x) = I_0(x) +I_B(x) \ ,
\label{app3::int_and_backg}
\end{equation}
where $I_0(x)$ is the intensity of the beam and $I_B(x)$ is the background intensity.
We assume also that the background signal is statistically independent from the beam intensity fluctuations.
For the background fluctuations we consider that
\begin{eqnarray}
\langle I_B(x) \rangle & = & C\ ,\\
\langle I_B(x_1) I_B(x_2) \rangle & = & C^2(1+\delta_{x_1,x_2}) \ ,
\label{app3::backs}
\end{eqnarray}
where $C<<max(I_0)$ and the background signal is not significant in the center of the beam.

Substituting Eq.\eqref{app3::int_and_backg} in the general expression for the 2-nd order intensity correlation function
\begin{equation}
g^{(2)}(x_1, x_2) = \frac{\langle I(x_1) I(x_2) \rangle}{\langle I(x_1)\rangle \langle I(x_2) \rangle}
\label{app3::g_2}
\end{equation}
we will determine how additional background could influence measured values of the 2-nd order intensity correlation function.

The same procedure can be done for the beams in the presence of jitter described in the previous section.
In this case correlation functions in the presence of jitter will be modified according to
\begin{eqnarray}
\langle I(x) \rangle_{jit} & = & p(x)*\langle I(x)\rangle\ ,\\
\langle I(x_1) I(x_2) \rangle_{jit} & = & p(x_1)\delta(x_2-x_1)*\langle I(x_1)I(x_2)\rangle\ .
\label{app3::jit}
\end{eqnarray}
Substituting here values of intensity $I(x)$ in the presence of noise as given in Eq.~\eqref{app3::int_and_backg} we will obtain final result.
This approach was used in simulations of the 2-nd order correlation function in the presence of jitter and background noise presented in Fig.~13(e,f).
%
%


\eject

\bibliography{references}

\begin{thebibliography}{64}
\expandafter\ifx\csname natexlab\endcsname\relax\def\natexlab#1{#1}\fi
\expandafter\ifx\csname bibnamefont\endcsname\relax
  \def\bibnamefont#1{#1}\fi
\expandafter\ifx\csname bibfnamefont\endcsname\relax
  \def\bibfnamefont#1{#1}\fi
\expandafter\ifx\csname citenamefont\endcsname\relax
  \def\citenamefont#1{#1}\fi
\expandafter\ifx\csname url\endcsname\relax
  \def\url#1{\texttt{#1}}\fi
\expandafter\ifx\csname urlprefix\endcsname\relax\def\urlprefix{URL }\fi
\providecommand{\bibinfo}[2]{#2}
\providecommand{\eprint}[2][]{\url{#2}}

\bibitem[{\citenamefont{Emma et~al.}(2010)\citenamefont{Emma, Akre, Arthur,
  Bionta, Bostedt, Bozek, Brachmann, Bucksbaum, Coffee, Decker et~al.}}]{LCLS}
\bibinfo{author}{\bibfnamefont{P.}~\bibnamefont{Emma}},
  \bibinfo{author}{\bibfnamefont{R.}~\bibnamefont{Akre}},
  \bibinfo{author}{\bibfnamefont{J.}~\bibnamefont{Arthur}},
  \bibinfo{author}{\bibfnamefont{R.}~\bibnamefont{Bionta}},
  \bibinfo{author}{\bibfnamefont{C.}~\bibnamefont{Bostedt}},
  \bibinfo{author}{\bibfnamefont{J.}~\bibnamefont{Bozek}},
  \bibinfo{author}{\bibfnamefont{A.}~\bibnamefont{Brachmann}},
  \bibinfo{author}{\bibfnamefont{P.}~\bibnamefont{Bucksbaum}},
  \bibinfo{author}{\bibfnamefont{R.}~\bibnamefont{Coffee}},
  \bibinfo{author}{\bibfnamefont{F.-J.} \bibnamefont{Decker}},
  \bibnamefont{et~al.}, \bibinfo{journal}{Nat. Photon.}
  \textbf{\bibinfo{volume}{4}}, \bibinfo{pages}{641} (\bibinfo{year}{2010}).

\bibitem[{\citenamefont{Ishikawa et~al.}(2012)\citenamefont{Ishikawa, Aoyagi,
  Asaka, Asano, Azumi, Bizen, Ego, Fukami, Fukui, Furukawa et~al.}}]{SCSS}
\bibinfo{author}{\bibfnamefont{T.}~\bibnamefont{Ishikawa}},
  \bibinfo{author}{\bibfnamefont{H.}~\bibnamefont{Aoyagi}},
  \bibinfo{author}{\bibfnamefont{T.}~\bibnamefont{Asaka}},
  \bibinfo{author}{\bibfnamefont{Y.}~\bibnamefont{Asano}},
  \bibinfo{author}{\bibfnamefont{N.}~\bibnamefont{Azumi}},
  \bibinfo{author}{\bibfnamefont{T.}~\bibnamefont{Bizen}},
  \bibinfo{author}{\bibfnamefont{H.}~\bibnamefont{Ego}},
  \bibinfo{author}{\bibfnamefont{K.}~\bibnamefont{Fukami}},
  \bibinfo{author}{\bibfnamefont{T.}~\bibnamefont{Fukui}},
  \bibinfo{author}{\bibfnamefont{Y.}~\bibnamefont{Furukawa}},
  \bibnamefont{et~al.}, \bibinfo{journal}{Nat. Photon.}
  \textbf{\bibinfo{volume}{6}}, \bibinfo{pages}{540} (\bibinfo{year}{2012}).

\bibitem[{\citenamefont{Altarelli et~al.}(2006)\citenamefont{Altarelli,
  Brinkmann, Chergui, Decking, Dobson, Dusterer, Grubel, Graeff, Graafsma,
  Hajdu et~al.}}]{XFEL}
\bibinfo{author}{\bibfnamefont{M.}~\bibnamefont{Altarelli}},
  \bibinfo{author}{\bibfnamefont{R.}~\bibnamefont{Brinkmann}},
  \bibinfo{author}{\bibfnamefont{M.}~\bibnamefont{Chergui}},
  \bibinfo{author}{\bibfnamefont{W.}~\bibnamefont{Decking}},
  \bibinfo{author}{\bibfnamefont{B.}~\bibnamefont{Dobson}},
  \bibinfo{author}{\bibfnamefont{S.}~\bibnamefont{Dusterer}},
  \bibinfo{author}{\bibfnamefont{G.}~\bibnamefont{Grubel}},
  \bibinfo{author}{\bibfnamefont{W.}~\bibnamefont{Graeff}},
  \bibinfo{author}{\bibfnamefont{H.}~\bibnamefont{Graafsma}},
  \bibinfo{author}{\bibfnamefont{J.}~\bibnamefont{Hajdu}},
  \bibnamefont{et~al.}, \bibinfo{type}{Tech. Rep.} \bibinfo{number}{DESY
  2006-097}, \bibinfo{institution}{European XFEL Project Team, Deutsches
  Elektronen-Synchrotron}, \bibinfo{address}{Hamburg} (\bibinfo{year}{2006}).

\bibitem[{\citenamefont{Seibert et~al.}(2011)\citenamefont{Seibert, Ekeberg,
  Maia, Svenda, Andreasson, J\"{o}nsson, Odi\'{c}, Iwan, Rocker, Westphal
  et~al.}}]{Seibert2011}
\bibinfo{author}{\bibfnamefont{M.~M.} \bibnamefont{Seibert}},
  \bibinfo{author}{\bibfnamefont{T.}~\bibnamefont{Ekeberg}},
  \bibinfo{author}{\bibfnamefont{F.~R. N.~C.} \bibnamefont{Maia}},
  \bibinfo{author}{\bibfnamefont{M.}~\bibnamefont{Svenda}},
  \bibinfo{author}{\bibfnamefont{J.}~\bibnamefont{Andreasson}},
  \bibinfo{author}{\bibfnamefont{O.}~\bibnamefont{J\"{o}nsson}},
  \bibinfo{author}{\bibfnamefont{D.}~\bibnamefont{Odi\'{c}}},
  \bibinfo{author}{\bibfnamefont{B.}~\bibnamefont{Iwan}},
  \bibinfo{author}{\bibfnamefont{A.}~\bibnamefont{Rocker}},
  \bibinfo{author}{\bibfnamefont{D.}~\bibnamefont{Westphal}},
  \bibnamefont{et~al.}, \bibinfo{journal}{Nature}
  \textbf{\bibinfo{volume}{470}}, \bibinfo{pages}{78} (\bibinfo{year}{2011}).

\bibitem[{\citenamefont{Chapman et~al.}(2011)\citenamefont{Chapman, Fromme,
  Barty, White, Kirian, Aquila, Hunter, Schulz, DePonte, Weierstall
  et~al.}}]{Chapman2011}
\bibinfo{author}{\bibfnamefont{H.~N.} \bibnamefont{Chapman}},
  \bibinfo{author}{\bibfnamefont{P.}~\bibnamefont{Fromme}},
  \bibinfo{author}{\bibfnamefont{A.}~\bibnamefont{Barty}},
  \bibinfo{author}{\bibfnamefont{T.~A.} \bibnamefont{White}},
  \bibinfo{author}{\bibfnamefont{R.~A.} \bibnamefont{Kirian}},
  \bibinfo{author}{\bibfnamefont{A.}~\bibnamefont{Aquila}},
  \bibinfo{author}{\bibfnamefont{M.~S.} \bibnamefont{Hunter}},
  \bibinfo{author}{\bibfnamefont{J.}~\bibnamefont{Schulz}},
  \bibinfo{author}{\bibfnamefont{D.~P.} \bibnamefont{DePonte}},
  \bibinfo{author}{\bibfnamefont{U.}~\bibnamefont{Weierstall}},
  \bibnamefont{et~al.}, \bibinfo{journal}{Nature}
  \textbf{\bibinfo{volume}{470}}, \bibinfo{pages}{73} (\bibinfo{year}{2011}).

\bibitem[{\citenamefont{Vinko et~al.}(2012)\citenamefont{Vinko, Ciricosta, Cho,
  Engelhorn, Chung, Brown, Burian, Chalupsk{\'{y}}, Falcone, Graves
  et~al.}}]{Vinko2012}
\bibinfo{author}{\bibfnamefont{S.~M.} \bibnamefont{Vinko}},
  \bibinfo{author}{\bibfnamefont{O.}~\bibnamefont{Ciricosta}},
  \bibinfo{author}{\bibfnamefont{B.~I.} \bibnamefont{Cho}},
  \bibinfo{author}{\bibfnamefont{K.}~\bibnamefont{Engelhorn}},
  \bibinfo{author}{\bibfnamefont{H.-K.} \bibnamefont{Chung}},
  \bibinfo{author}{\bibfnamefont{C.~R.~D.} \bibnamefont{Brown}},
  \bibinfo{author}{\bibfnamefont{T.}~\bibnamefont{Burian}},
  \bibinfo{author}{\bibfnamefont{J.}~\bibnamefont{Chalupsk{\'{y}}}},
  \bibinfo{author}{\bibfnamefont{R.~W.} \bibnamefont{Falcone}},
  \bibinfo{author}{\bibfnamefont{C.}~\bibnamefont{Graves}},
  \bibnamefont{et~al.}, \bibinfo{journal}{Nature}
  \textbf{\bibinfo{volume}{482}}, \bibinfo{pages}{59} (\bibinfo{year}{2012}).

\bibitem[{\citenamefont{Schropp et~al.}(2015)\citenamefont{Schropp, Hoppe,
  Meier, Patommel, Seiboth, Ping, Hicks, Beckwith, Collins, Higginbotham
  et~al.}}]{Schropp2015}
\bibinfo{author}{\bibfnamefont{A.}~\bibnamefont{Schropp}},
  \bibinfo{author}{\bibfnamefont{R.}~\bibnamefont{Hoppe}},
  \bibinfo{author}{\bibfnamefont{V.}~\bibnamefont{Meier}},
  \bibinfo{author}{\bibfnamefont{J.}~\bibnamefont{Patommel}},
  \bibinfo{author}{\bibfnamefont{F.}~\bibnamefont{Seiboth}},
  \bibinfo{author}{\bibfnamefont{Y.}~\bibnamefont{Ping}},
  \bibinfo{author}{\bibfnamefont{D.~G.} \bibnamefont{Hicks}},
  \bibinfo{author}{\bibfnamefont{M.~A.} \bibnamefont{Beckwith}},
  \bibinfo{author}{\bibfnamefont{G.~W.} \bibnamefont{Collins}},
  \bibinfo{author}{\bibfnamefont{A.}~\bibnamefont{Higginbotham}},
  \bibnamefont{et~al.}, \bibinfo{journal}{Sci. Rep.}
  \textbf{\bibinfo{volume}{5}}, \bibinfo{pages}{11089} (\bibinfo{year}{2015}).

\bibitem[{\citenamefont{Liekhus-Schmaltz
  et~al.}(2015)\citenamefont{Liekhus-Schmaltz, Tenney, Osipov,
  Sanchez-Gonzalez, Berrah, Boll, Bomme, Bostedt, Bozek, Carron
  et~al.}}]{LiekhusSchmaltz2015}
\bibinfo{author}{\bibfnamefont{C.~E.} \bibnamefont{Liekhus-Schmaltz}},
  \bibinfo{author}{\bibfnamefont{I.}~\bibnamefont{Tenney}},
  \bibinfo{author}{\bibfnamefont{T.}~\bibnamefont{Osipov}},
  \bibinfo{author}{\bibfnamefont{A.}~\bibnamefont{Sanchez-Gonzalez}},
  \bibinfo{author}{\bibfnamefont{N.}~\bibnamefont{Berrah}},
  \bibinfo{author}{\bibfnamefont{R.}~\bibnamefont{Boll}},
  \bibinfo{author}{\bibfnamefont{C.}~\bibnamefont{Bomme}},
  \bibinfo{author}{\bibfnamefont{C.}~\bibnamefont{Bostedt}},
  \bibinfo{author}{\bibfnamefont{J.~D.} \bibnamefont{Bozek}},
  \bibinfo{author}{\bibfnamefont{S.}~\bibnamefont{Carron}},
  \bibnamefont{et~al.}, \bibinfo{journal}{Nat. Commun.}
  \textbf{\bibinfo{volume}{6}}, \bibinfo{pages}{8199} (\bibinfo{year}{2015}).

\bibitem[{\citenamefont{Young et~al.}(2010)\citenamefont{Young, Kanter,
  Kr{\"{a}}ssig, Li, March, Pratt, Santra, Southworth, Rohringer, DiMauro
  et~al.}}]{Young2010}
\bibinfo{author}{\bibfnamefont{L.}~\bibnamefont{Young}},
  \bibinfo{author}{\bibfnamefont{E.~P.} \bibnamefont{Kanter}},
  \bibinfo{author}{\bibfnamefont{B.}~\bibnamefont{Kr{\"{a}}ssig}},
  \bibinfo{author}{\bibfnamefont{Y.}~\bibnamefont{Li}},
  \bibinfo{author}{\bibfnamefont{A.~M.} \bibnamefont{March}},
  \bibinfo{author}{\bibfnamefont{S.~T.} \bibnamefont{Pratt}},
  \bibinfo{author}{\bibfnamefont{R.}~\bibnamefont{Santra}},
  \bibinfo{author}{\bibfnamefont{S.~H.} \bibnamefont{Southworth}},
  \bibinfo{author}{\bibfnamefont{N.}~\bibnamefont{Rohringer}},
  \bibinfo{author}{\bibfnamefont{L.~F.} \bibnamefont{DiMauro}},
  \bibnamefont{et~al.}, \bibinfo{journal}{Nature}
  \textbf{\bibinfo{volume}{466}}, \bibinfo{pages}{56} (\bibinfo{year}{2010}).

\bibitem[{\citenamefont{Trigo et~al.}(2013)\citenamefont{Trigo, Fuchs, Chen,
  Jiang, Cammarata, Fahy, Fritz, Gaffney, Ghimire, Higginbotham
  et~al.}}]{Trigo2013}
\bibinfo{author}{\bibfnamefont{M.}~\bibnamefont{Trigo}},
  \bibinfo{author}{\bibfnamefont{M.}~\bibnamefont{Fuchs}},
  \bibinfo{author}{\bibfnamefont{J.}~\bibnamefont{Chen}},
  \bibinfo{author}{\bibfnamefont{M.~P.} \bibnamefont{Jiang}},
  \bibinfo{author}{\bibfnamefont{M.}~\bibnamefont{Cammarata}},
  \bibinfo{author}{\bibfnamefont{S.}~\bibnamefont{Fahy}},
  \bibinfo{author}{\bibfnamefont{D.~M.} \bibnamefont{Fritz}},
  \bibinfo{author}{\bibfnamefont{K.}~\bibnamefont{Gaffney}},
  \bibinfo{author}{\bibfnamefont{S.}~\bibnamefont{Ghimire}},
  \bibinfo{author}{\bibfnamefont{A.}~\bibnamefont{Higginbotham}},
  \bibnamefont{et~al.}, \bibinfo{journal}{Nat. Phys.}
  \textbf{\bibinfo{volume}{9}}, \bibinfo{pages}{790} (\bibinfo{year}{2013}).

\bibitem[{\citenamefont{Dronyak et~al.}(2012)\citenamefont{Dronyak, Gulden,
  Yefanov, Singer, Gorniak, Senkbeil, Meijer, Al-Shemmary, Hallmann, Mai
  et~al.}}]{Dronyak2012}
\bibinfo{author}{\bibfnamefont{R.}~\bibnamefont{Dronyak}},
  \bibinfo{author}{\bibfnamefont{J.}~\bibnamefont{Gulden}},
  \bibinfo{author}{\bibfnamefont{O.~M.} \bibnamefont{Yefanov}},
  \bibinfo{author}{\bibfnamefont{A.}~\bibnamefont{Singer}},
  \bibinfo{author}{\bibfnamefont{T.}~\bibnamefont{Gorniak}},
  \bibinfo{author}{\bibfnamefont{T.}~\bibnamefont{Senkbeil}},
  \bibinfo{author}{\bibfnamefont{J.-M.} \bibnamefont{Meijer}},
  \bibinfo{author}{\bibfnamefont{A.}~\bibnamefont{Al-Shemmary}},
  \bibinfo{author}{\bibfnamefont{J.}~\bibnamefont{Hallmann}},
  \bibinfo{author}{\bibfnamefont{D.~D.} \bibnamefont{Mai}},
  \bibnamefont{et~al.}, \bibinfo{journal}{Phys. Rev. B}
  \textbf{\bibinfo{volume}{86}}, \bibinfo{pages}{064303}
  (\bibinfo{year}{2012}).

\bibitem[{\citenamefont{Clark et~al.}(2013)\citenamefont{Clark, Beitra, Xiong,
  Higginbotham, Fritz, Lemke, Zhu, Chollet, Williams, Messerschmidt
  et~al.}}]{Clark2013}
\bibinfo{author}{\bibfnamefont{J.~N.} \bibnamefont{Clark}},
  \bibinfo{author}{\bibfnamefont{L.}~\bibnamefont{Beitra}},
  \bibinfo{author}{\bibfnamefont{G.}~\bibnamefont{Xiong}},
  \bibinfo{author}{\bibfnamefont{A.}~\bibnamefont{Higginbotham}},
  \bibinfo{author}{\bibfnamefont{D.~M.} \bibnamefont{Fritz}},
  \bibinfo{author}{\bibfnamefont{H.~T.} \bibnamefont{Lemke}},
  \bibinfo{author}{\bibfnamefont{D.}~\bibnamefont{Zhu}},
  \bibinfo{author}{\bibfnamefont{M.}~\bibnamefont{Chollet}},
  \bibinfo{author}{\bibfnamefont{G.~J.} \bibnamefont{Williams}},
  \bibinfo{author}{\bibfnamefont{M.}~\bibnamefont{Messerschmidt}},
  \bibnamefont{et~al.}, \bibinfo{journal}{Science}
  \textbf{\bibinfo{volume}{341}}, \bibinfo{pages}{56} (\bibinfo{year}{2013}).

\bibitem[{\citenamefont{Singer et~al.}(2016)\citenamefont{Singer, Patel,
  Kukreja, Uhl\'{\i}\ifmmode~\check{r}\else \v{r}\fi{}, Wingert, Festersen,
  Zhu, Glownia, Lemke, Nelson et~al.}}]{Singer2016}
\bibinfo{author}{\bibfnamefont{A.}~\bibnamefont{Singer}},
  \bibinfo{author}{\bibfnamefont{S.~K.~K.} \bibnamefont{Patel}},
  \bibinfo{author}{\bibfnamefont{R.}~\bibnamefont{Kukreja}},
  \bibinfo{author}{\bibfnamefont{V.}~\bibnamefont{Uhl\'{\i}\ifmmode~\check{r}\else
  \v{r}\fi{}}}, \bibinfo{author}{\bibfnamefont{J.}~\bibnamefont{Wingert}},
  \bibinfo{author}{\bibfnamefont{S.}~\bibnamefont{Festersen}},
  \bibinfo{author}{\bibfnamefont{D.}~\bibnamefont{Zhu}},
  \bibinfo{author}{\bibfnamefont{J.~M.} \bibnamefont{Glownia}},
  \bibinfo{author}{\bibfnamefont{H.~T.} \bibnamefont{Lemke}},
  \bibinfo{author}{\bibfnamefont{S.}~\bibnamefont{Nelson}},
  \bibnamefont{et~al.}, \bibinfo{journal}{Phys. Rev. Lett.}
  \textbf{\bibinfo{volume}{117}}, \bibinfo{pages}{056401}
  (\bibinfo{year}{2016}).

\bibitem[{\citenamefont{Beaud et~al.}(2014)\citenamefont{Beaud, Caviezel,
  Mariager, Rettig, Ingold, Dornes, Huang, Johnson, Radovic, Huber
  et~al.}}]{Beaud2014}
\bibinfo{author}{\bibfnamefont{P.}~\bibnamefont{Beaud}},
  \bibinfo{author}{\bibfnamefont{A.}~\bibnamefont{Caviezel}},
  \bibinfo{author}{\bibfnamefont{S.~O.} \bibnamefont{Mariager}},
  \bibinfo{author}{\bibfnamefont{L.}~\bibnamefont{Rettig}},
  \bibinfo{author}{\bibfnamefont{G.}~\bibnamefont{Ingold}},
  \bibinfo{author}{\bibfnamefont{C.}~\bibnamefont{Dornes}},
  \bibinfo{author}{\bibfnamefont{S.-W.} \bibnamefont{Huang}},
  \bibinfo{author}{\bibfnamefont{J.~A.} \bibnamefont{Johnson}},
  \bibinfo{author}{\bibfnamefont{M.}~\bibnamefont{Radovic}},
  \bibinfo{author}{\bibfnamefont{T.}~\bibnamefont{Huber}},
  \bibnamefont{et~al.}, \bibinfo{journal}{Nat. Mater.}
  \textbf{\bibinfo{volume}{13}}, \bibinfo{pages}{923} (\bibinfo{year}{2014}).

\bibitem[{\citenamefont{Nugent}(2010)}]{Nugent2010}
\bibinfo{author}{\bibfnamefont{K.~A.} \bibnamefont{Nugent}},
  \bibinfo{journal}{Adv. Phys.} \textbf{\bibinfo{volume}{59}},
  \bibinfo{pages}{1} (\bibinfo{year}{2010}).

\bibitem[{\citenamefont{Chapman and Nugent}(2010)}]{Chapman2010}
\bibinfo{author}{\bibfnamefont{H.~N.} \bibnamefont{Chapman}} \bibnamefont{and}
  \bibinfo{author}{\bibfnamefont{K.~A.} \bibnamefont{Nugent}},
  \bibinfo{journal}{Nat. Phot.} \textbf{\bibinfo{volume}{4}},
  \bibinfo{pages}{833} (\bibinfo{year}{2010}).

\bibitem[{\citenamefont{Vartanyants and Yefanov}(2015)}]{Vartanyants2015}
\bibinfo{author}{\bibfnamefont{I.~A.} \bibnamefont{Vartanyants}}
  \bibnamefont{and} \bibinfo{author}{\bibfnamefont{O.~M.}
  \bibnamefont{Yefanov}}, in \emph{\bibinfo{booktitle}{X-Ray Diffraction.
  Modern Experimental Techniques}}, edited by
  \bibinfo{editor}{\bibfnamefont{O.~H.} \bibnamefont{Seeck}} \bibnamefont{and}
  \bibinfo{editor}{\bibfnamefont{B.~M.} \bibnamefont{Murphy}}
  (\bibinfo{publisher}{Pan Stanford Publishing}, \bibinfo{address}{Singapore},
  \bibinfo{year}{2015}), pp. \bibinfo{pages}{337--380}.

\bibitem[{\citenamefont{Vartanyants and Robinson}(2001)}]{Vartanyants2001}
\bibinfo{author}{\bibfnamefont{I.~A.} \bibnamefont{Vartanyants}}
  \bibnamefont{and} \bibinfo{author}{\bibfnamefont{I.~K.}
  \bibnamefont{Robinson}}, \bibinfo{journal}{J. Phys.: Cond. Matt.}
  \textbf{\bibinfo{volume}{13}}, \bibinfo{pages}{10593} (\bibinfo{year}{2001}).

\bibitem[{\citenamefont{Vartanyants and Robinson}(2003)}]{Vartanyants2003}
\bibinfo{author}{\bibfnamefont{I.~A.} \bibnamefont{Vartanyants}}
  \bibnamefont{and} \bibinfo{author}{\bibfnamefont{I.~K.}
  \bibnamefont{Robinson}}, \bibinfo{journal}{Opt. Comm.}
  \textbf{\bibinfo{volume}{222}}, \bibinfo{pages}{29} (\bibinfo{year}{2003}).

\bibitem[{\citenamefont{Williams et~al.}(2007)\citenamefont{Williams, Quiney,
  Peele, and Nugent}}]{Williams2007}
\bibinfo{author}{\bibfnamefont{G.~J.} \bibnamefont{Williams}},
  \bibinfo{author}{\bibfnamefont{H.~M.} \bibnamefont{Quiney}},
  \bibinfo{author}{\bibfnamefont{A.~G.} \bibnamefont{Peele}}, \bibnamefont{and}
  \bibinfo{author}{\bibfnamefont{K.~A.} \bibnamefont{Nugent}},
  \bibinfo{journal}{Phys. Rev. B} \textbf{\bibinfo{volume}{75}},
  \bibinfo{pages}{104102} (\bibinfo{year}{2007}).

\bibitem[{\citenamefont{Whitehead et~al.}(2009)\citenamefont{Whitehead,
  Williams, Quiney, Vine, Dilanian, Flewett, Nugent, Peele, Balaur, and
  McNulty}}]{Whitehead2009}
\bibinfo{author}{\bibfnamefont{L.~W.} \bibnamefont{Whitehead}},
  \bibinfo{author}{\bibfnamefont{G.~J.} \bibnamefont{Williams}},
  \bibinfo{author}{\bibfnamefont{H.~M.} \bibnamefont{Quiney}},
  \bibinfo{author}{\bibfnamefont{D.~J.} \bibnamefont{Vine}},
  \bibinfo{author}{\bibfnamefont{R.~A.} \bibnamefont{Dilanian}},
  \bibinfo{author}{\bibfnamefont{S.}~\bibnamefont{Flewett}},
  \bibinfo{author}{\bibfnamefont{K.~A.} \bibnamefont{Nugent}},
  \bibinfo{author}{\bibfnamefont{A.~G.} \bibnamefont{Peele}},
  \bibinfo{author}{\bibfnamefont{E.}~\bibnamefont{Balaur}}, \bibnamefont{and}
  \bibinfo{author}{\bibfnamefont{I.}~\bibnamefont{McNulty}},
  \bibinfo{journal}{Phys. Rev. Lett.} \textbf{\bibinfo{volume}{103}},
  \bibinfo{pages}{243902} (\bibinfo{year}{2009}).

\bibitem[{\citenamefont{Saldin et~al.}(2000)\citenamefont{Saldin,
  Schneidmiller, and Yurkov}}]{SaldinBook}
\bibinfo{author}{\bibfnamefont{E.~L.} \bibnamefont{Saldin}},
  \bibinfo{author}{\bibfnamefont{E.~A.} \bibnamefont{Schneidmiller}},
  \bibnamefont{and} \bibinfo{author}{\bibfnamefont{M.~V.}
  \bibnamefont{Yurkov}}, \emph{\bibinfo{title}{The Physics of Free Electron
  Laser}} (\bibinfo{publisher}{Springer-Verlag}, \bibinfo{address}{Berlin},
  \bibinfo{year}{2000}).

\bibitem[{\citenamefont{Saldin et~al.}(2006)\citenamefont{Saldin,
  Schneidmiller, and Yurkov}}]{Saldin2006}
\bibinfo{author}{\bibfnamefont{E.~L.} \bibnamefont{Saldin}},
  \bibinfo{author}{\bibfnamefont{E.~A.} \bibnamefont{Schneidmiller}},
  \bibnamefont{and} \bibinfo{author}{\bibfnamefont{M.~V.}
  \bibnamefont{Yurkov}}, \bibinfo{journal}{Phys. Rev. ST Accel. Beams}
  \textbf{\bibinfo{volume}{9}}, \bibinfo{pages}{050702} (\bibinfo{year}{2006}).

\bibitem[{\citenamefont{Vartanyants et~al.}(2011)\citenamefont{Vartanyants,
  Singer, Mancuso, Yefanov, Sakdinawat, Liu, Bang, Williams, Cadenazzi, Abbey
  et~al.}}]{Vartanyants2011}
\bibinfo{author}{\bibfnamefont{I.~A.} \bibnamefont{Vartanyants}},
  \bibinfo{author}{\bibfnamefont{A.}~\bibnamefont{Singer}},
  \bibinfo{author}{\bibfnamefont{A.~P.} \bibnamefont{Mancuso}},
  \bibinfo{author}{\bibfnamefont{O.~M.} \bibnamefont{Yefanov}},
  \bibinfo{author}{\bibfnamefont{A.}~\bibnamefont{Sakdinawat}},
  \bibinfo{author}{\bibfnamefont{Y.}~\bibnamefont{Liu}},
  \bibinfo{author}{\bibfnamefont{E.}~\bibnamefont{Bang}},
  \bibinfo{author}{\bibfnamefont{G.~J.} \bibnamefont{Williams}},
  \bibinfo{author}{\bibfnamefont{G.}~\bibnamefont{Cadenazzi}},
  \bibinfo{author}{\bibfnamefont{B.}~\bibnamefont{Abbey}},
  \bibnamefont{et~al.}, \bibinfo{journal}{Phys. Rev. Lett.}
  \textbf{\bibinfo{volume}{107}}, \bibinfo{pages}{144801}
  (\bibinfo{year}{2011}).

\bibitem[{\citenamefont{Singer et~al.}(2012)\citenamefont{Singer, Sorgenfrei,
  Mancuso, Gerasimova, Yefanov, Gulden, Gorniak, Senkbeil, Sakdinawat, Liu
  et~al.}}]{Singer2012}
\bibinfo{author}{\bibfnamefont{A.}~\bibnamefont{Singer}},
  \bibinfo{author}{\bibfnamefont{F.}~\bibnamefont{Sorgenfrei}},
  \bibinfo{author}{\bibfnamefont{A.}~\bibnamefont{Mancuso}},
  \bibinfo{author}{\bibfnamefont{N.}~\bibnamefont{Gerasimova}},
  \bibinfo{author}{\bibfnamefont{O.}~\bibnamefont{Yefanov}},
  \bibinfo{author}{\bibfnamefont{J.}~\bibnamefont{Gulden}},
  \bibinfo{author}{\bibfnamefont{T.}~\bibnamefont{Gorniak}},
  \bibinfo{author}{\bibfnamefont{T.}~\bibnamefont{Senkbeil}},
  \bibinfo{author}{\bibfnamefont{A.}~\bibnamefont{Sakdinawat}},
  \bibinfo{author}{\bibfnamefont{Y.}~\bibnamefont{Liu}}, \bibnamefont{et~al.},
  \bibinfo{journal}{Opt. Express} \textbf{\bibinfo{volume}{20}},
  \bibinfo{pages}{17480} (\bibinfo{year}{2012}).

\bibitem[{\citenamefont{Skopintsev et~al.}(2014)\citenamefont{Skopintsev,
  Singer, Bach, Müller, Beyersdorff, Schleitzer, Gorobtsov, Shabalin, Kurta,
  Dzhigaev et~al.}}]{Skopintsev2014}
\bibinfo{author}{\bibfnamefont{P.}~\bibnamefont{Skopintsev}},
  \bibinfo{author}{\bibfnamefont{A.}~\bibnamefont{Singer}},
  \bibinfo{author}{\bibfnamefont{J.}~\bibnamefont{Bach}},
  \bibinfo{author}{\bibfnamefont{L.}~\bibnamefont{Müller}},
  \bibinfo{author}{\bibfnamefont{B.}~\bibnamefont{Beyersdorff}},
  \bibinfo{author}{\bibfnamefont{S.}~\bibnamefont{Schleitzer}},
  \bibinfo{author}{\bibfnamefont{O.}~\bibnamefont{Gorobtsov}},
  \bibinfo{author}{\bibfnamefont{A.}~\bibnamefont{Shabalin}},
  \bibinfo{author}{\bibfnamefont{R.~P.} \bibnamefont{Kurta}},
  \bibinfo{author}{\bibfnamefont{D.}~\bibnamefont{Dzhigaev}},
  \bibnamefont{et~al.}, \bibinfo{journal}{J. Synch. Rad.}
  \textbf{\bibinfo{volume}{21}}, \bibinfo{pages}{722} (\bibinfo{year}{2014}).

\bibitem[{\citenamefont{Roling et~al.}(2011)\citenamefont{Roling, Siemer,
  W{\"{o}}stmann, Zacharias, Mitzner, Singer, Tiedtke, and
  Vartanyants}}]{Roling2011}
\bibinfo{author}{\bibfnamefont{S.}~\bibnamefont{Roling}},
  \bibinfo{author}{\bibfnamefont{B.}~\bibnamefont{Siemer}},
  \bibinfo{author}{\bibfnamefont{M.}~\bibnamefont{W{\"{o}}stmann}},
  \bibinfo{author}{\bibfnamefont{H.}~\bibnamefont{Zacharias}},
  \bibinfo{author}{\bibfnamefont{R.}~\bibnamefont{Mitzner}},
  \bibinfo{author}{\bibfnamefont{A.}~\bibnamefont{Singer}},
  \bibinfo{author}{\bibfnamefont{K.}~\bibnamefont{Tiedtke}}, \bibnamefont{and}
  \bibinfo{author}{\bibfnamefont{I.~A.} \bibnamefont{Vartanyants}},
  \bibinfo{journal}{Phys. Rev. ST Accel. Beams} \textbf{\bibinfo{volume}{14}},
  \bibinfo{pages}{080701} (\bibinfo{year}{2011}).

\bibitem[{\citenamefont{Hilbert et~al.}(2014)\citenamefont{Hilbert,
  R{\"{o}}del, Brenner, D{\"{o}}ppner, D{\"{u}}sterer, Dziarzhytski, Fletcher,
  F{\"{o}}rster, Glenzer, Harmand et~al.}}]{Hilbert2014}
\bibinfo{author}{\bibfnamefont{V.}~\bibnamefont{Hilbert}},
  \bibinfo{author}{\bibfnamefont{C.}~\bibnamefont{R{\"{o}}del}},
  \bibinfo{author}{\bibfnamefont{G.}~\bibnamefont{Brenner}},
  \bibinfo{author}{\bibfnamefont{T.}~\bibnamefont{D{\"{o}}ppner}},
  \bibinfo{author}{\bibfnamefont{S.}~\bibnamefont{D{\"{u}}sterer}},
  \bibinfo{author}{\bibfnamefont{S.}~\bibnamefont{Dziarzhytski}},
  \bibinfo{author}{\bibfnamefont{L.}~\bibnamefont{Fletcher}},
  \bibinfo{author}{\bibfnamefont{E.}~\bibnamefont{F{\"{o}}rster}},
  \bibinfo{author}{\bibfnamefont{S.~H.} \bibnamefont{Glenzer}},
  \bibinfo{author}{\bibfnamefont{M.}~\bibnamefont{Harmand}},
  \bibnamefont{et~al.}, \bibinfo{journal}{Appl. Phys. Lett.}
  \textbf{\bibinfo{volume}{105}}, \bibinfo{pages}{101102}
  (\bibinfo{year}{2014}).

\bibitem[{\citenamefont{Singer et~al.}(2013)\citenamefont{Singer, Lorenz,
  Sorgenfrei, Gerasimova, Gulden, Yefanov, Kurta, Shabalin, Dronyak, Treusch
  et~al.}}]{Singer2013}
\bibinfo{author}{\bibfnamefont{A.}~\bibnamefont{Singer}},
  \bibinfo{author}{\bibfnamefont{U.}~\bibnamefont{Lorenz}},
  \bibinfo{author}{\bibfnamefont{F.}~\bibnamefont{Sorgenfrei}},
  \bibinfo{author}{\bibfnamefont{N.}~\bibnamefont{Gerasimova}},
  \bibinfo{author}{\bibfnamefont{J.}~\bibnamefont{Gulden}},
  \bibinfo{author}{\bibfnamefont{O.~M.} \bibnamefont{Yefanov}},
  \bibinfo{author}{\bibfnamefont{R.~P.} \bibnamefont{Kurta}},
  \bibinfo{author}{\bibfnamefont{A.}~\bibnamefont{Shabalin}},
  \bibinfo{author}{\bibfnamefont{R.}~\bibnamefont{Dronyak}},
  \bibinfo{author}{\bibfnamefont{R.}~\bibnamefont{Treusch}},
  \bibnamefont{et~al.}, \bibinfo{journal}{Phys. Rev. Lett.}
  \textbf{\bibinfo{volume}{111}}, \bibinfo{pages}{034802}
  (\bibinfo{year}{2013}).

\bibitem[{\citenamefont{Vartanyants and Singer}(2015)}]{Vartaniants2015a}
\bibinfo{author}{\bibfnamefont{I.~A.} \bibnamefont{Vartanyants}}
  \bibnamefont{and} \bibinfo{author}{\bibfnamefont{A.}~\bibnamefont{Singer}},
  in \emph{\bibinfo{booktitle}{Synchrotron Light Sources and Free-Electron
  Lasers}}, edited by
  \bibinfo{editor}{\bibfnamefont{E.}~\bibnamefont{Jaeschke}},
  \bibinfo{editor}{\bibfnamefont{S.}~\bibnamefont{Khan}},
  \bibinfo{editor}{\bibfnamefont{J.}~\bibnamefont{Schneider}},
  \bibnamefont{and} \bibinfo{editor}{\bibfnamefont{J.}~\bibnamefont{Hastings}}
  (\bibinfo{publisher}{Springer}, \bibinfo{address}{Switzerland},
  \bibinfo{year}{2015}), pp. \bibinfo{pages}{1--38}.

\bibitem[{\citenamefont{{Hanbury Brown} and Twiss}(1956)}]{Hanbury}
\bibinfo{author}{\bibfnamefont{R.}~\bibnamefont{{Hanbury Brown}}}
  \bibnamefont{and} \bibinfo{author}{\bibfnamefont{R.~Q.} \bibnamefont{Twiss}},
  \bibinfo{journal}{Nature} \textbf{\bibinfo{volume}{177}}, \bibinfo{pages}{27}
  (\bibinfo{year}{1956}).

\bibitem[{\citenamefont{Twiss et~al.}(1957)\citenamefont{Twiss, Little, and
  Hanbury~Brown}}]{Twiss}
\bibinfo{author}{\bibfnamefont{R.~Q.} \bibnamefont{Twiss}},
  \bibinfo{author}{\bibfnamefont{A.~G.~G.} \bibnamefont{Little}},
  \bibnamefont{and}
  \bibinfo{author}{\bibfnamefont{R.}~\bibnamefont{Hanbury~Brown}},
  \bibinfo{journal}{Nature} \textbf{\bibinfo{volume}{180}},
  \bibinfo{pages}{324} (\bibinfo{year}{1957}).

\bibitem[{\citenamefont{Glauber}(1963)}]{Glauber1963}
\bibinfo{author}{\bibfnamefont{R.~J.} \bibnamefont{Glauber}},
  \bibinfo{journal}{Phys. Rev.} \textbf{\bibinfo{volume}{130}},
  \bibinfo{pages}{2529} (\bibinfo{year}{1963}).

\bibitem[{\citenamefont{Schellekens et~al.}(2005)\citenamefont{Schellekens,
  Hoppeler, Perrin, Viana~Gomes, Boiron, Aspect, and
  Westbrook}}]{Schellekens2005}
\bibinfo{author}{\bibfnamefont{M.}~\bibnamefont{Schellekens}},
  \bibinfo{author}{\bibfnamefont{R.}~\bibnamefont{Hoppeler}},
  \bibinfo{author}{\bibfnamefont{A.}~\bibnamefont{Perrin}},
  \bibinfo{author}{\bibfnamefont{J.}~\bibnamefont{Viana~Gomes}},
  \bibinfo{author}{\bibfnamefont{D.}~\bibnamefont{Boiron}},
  \bibinfo{author}{\bibfnamefont{A.}~\bibnamefont{Aspect}}, \bibnamefont{and}
  \bibinfo{author}{\bibfnamefont{C.~I.} \bibnamefont{Westbrook}},
  \bibinfo{journal}{Science} \textbf{\bibinfo{volume}{310}},
  \bibinfo{pages}{648} (\bibinfo{year}{2005}).

\bibitem[{\citenamefont{Baym}(1998)}]{Baym1998}
\bibinfo{author}{\bibfnamefont{G.}~\bibnamefont{Baym}}, \bibinfo{journal}{Acta
  Phys. Pol. B} \textbf{\bibinfo{volume}{29}} (\bibinfo{year}{1998}).

\bibitem[{\citenamefont{Goldberger et~al.}(1966)\citenamefont{Goldberger,
  Lewis, and Watson}}]{Goldberger}
\bibinfo{author}{\bibfnamefont{M.~L.} \bibnamefont{Goldberger}},
  \bibinfo{author}{\bibfnamefont{H.~W.} \bibnamefont{Lewis}}, \bibnamefont{and}
  \bibinfo{author}{\bibfnamefont{K.~M.} \bibnamefont{Watson}},
  \bibinfo{journal}{Phys. Rev.} \textbf{\bibinfo{volume}{142}},
  \bibinfo{pages}{25} (\bibinfo{year}{1966}).

\bibitem[{\citenamefont{Gluskin}(1991)}]{Gluskin1991}
\bibinfo{author}{\bibfnamefont{E.}~\bibnamefont{Gluskin}},
  \bibinfo{journal}{Proc. of PAC} \textbf{\bibinfo{volume}{2}},
  \bibinfo{pages}{1169} (\bibinfo{year}{1991}).

\bibitem[{\citenamefont{Ikonen}(1992)}]{Ikonen1992}
\bibinfo{author}{\bibfnamefont{E.}~\bibnamefont{Ikonen}},
  \bibinfo{journal}{Phys. Rev. Lett.} \textbf{\bibinfo{volume}{68}},
  \bibinfo{pages}{2759} (\bibinfo{year}{1992}).

\bibitem[{\citenamefont{Gluskin et~al.}(1999)\citenamefont{Gluskin, Alp,
  McNulty, Sturhahn, and Sutter}}]{Gluskin1999}
\bibinfo{author}{\bibfnamefont{E.}~\bibnamefont{Gluskin}},
  \bibinfo{author}{\bibfnamefont{E.~E.} \bibnamefont{Alp}},
  \bibinfo{author}{\bibfnamefont{I.}~\bibnamefont{McNulty}},
  \bibinfo{author}{\bibfnamefont{W.}~\bibnamefont{Sturhahn}}, \bibnamefont{and}
  \bibinfo{author}{\bibfnamefont{J.}~\bibnamefont{Sutter}},
  \bibinfo{journal}{J. Synch. Rad.} \textbf{\bibinfo{volume}{6}},
  \bibinfo{pages}{1065} (\bibinfo{year}{1999}).

\bibitem[{\citenamefont{Yabashi et~al.}(2001)\citenamefont{Yabashi, Tamasaku,
  and Ishikawa}}]{Yabashi2001}
\bibinfo{author}{\bibfnamefont{M.}~\bibnamefont{Yabashi}},
  \bibinfo{author}{\bibfnamefont{K.}~\bibnamefont{Tamasaku}}, \bibnamefont{and}
  \bibinfo{author}{\bibfnamefont{T.}~\bibnamefont{Ishikawa}},
  \bibinfo{journal}{Phys. Rev. Lett.} \textbf{\bibinfo{volume}{87}},
  \bibinfo{pages}{140801} (\bibinfo{year}{2001}).

\bibitem[{\citenamefont{Yabashi et~al.}(2002)\citenamefont{Yabashi, Tamasaku,
  and Ishikawa}}]{Yabashi2002}
\bibinfo{author}{\bibfnamefont{M.}~\bibnamefont{Yabashi}},
  \bibinfo{author}{\bibfnamefont{K.}~\bibnamefont{Tamasaku}}, \bibnamefont{and}
  \bibinfo{author}{\bibfnamefont{T.}~\bibnamefont{Ishikawa}},
  \bibinfo{journal}{Phys. Rev. Lett.} \textbf{\bibinfo{volume}{88}},
  \bibinfo{pages}{244801} (\bibinfo{year}{2002}).

\bibitem[{\citenamefont{Singer et~al.}(2014)\citenamefont{Singer, Lorenz,
  Marras, Klyuev, Becker, Schlage, Skopintsev, Gorobtsov, Shabalin, Wille
  et~al.}}]{Singer2014}
\bibinfo{author}{\bibfnamefont{A.}~\bibnamefont{Singer}},
  \bibinfo{author}{\bibfnamefont{U.}~\bibnamefont{Lorenz}},
  \bibinfo{author}{\bibfnamefont{A.}~\bibnamefont{Marras}},
  \bibinfo{author}{\bibfnamefont{A.}~\bibnamefont{Klyuev}},
  \bibinfo{author}{\bibfnamefont{J.}~\bibnamefont{Becker}},
  \bibinfo{author}{\bibfnamefont{K.}~\bibnamefont{Schlage}},
  \bibinfo{author}{\bibfnamefont{P.}~\bibnamefont{Skopintsev}},
  \bibinfo{author}{\bibfnamefont{O.}~\bibnamefont{Gorobtsov}},
  \bibinfo{author}{\bibfnamefont{A.}~\bibnamefont{Shabalin}},
  \bibinfo{author}{\bibfnamefont{H.-C.} \bibnamefont{Wille}},
  \bibnamefont{et~al.}, \bibinfo{journal}{Phys. Rev. Lett.}
  \textbf{\bibinfo{volume}{113}}, \bibinfo{pages}{064801}
  (\bibinfo{year}{2014}).

\bibitem[{\citenamefont{Song et~al.}(2014)\citenamefont{Song, Zhu, Singer, Wu,
  Sikorski, Chollet, Lemke, Alonso-Mori, Glownia, Krzywinski
  et~al.}}]{Song2014}
\bibinfo{author}{\bibfnamefont{S.}~\bibnamefont{Song}},
  \bibinfo{author}{\bibfnamefont{D.}~\bibnamefont{Zhu}},
  \bibinfo{author}{\bibfnamefont{A.}~\bibnamefont{Singer}},
  \bibinfo{author}{\bibfnamefont{J.}~\bibnamefont{Wu}},
  \bibinfo{author}{\bibfnamefont{M.}~\bibnamefont{Sikorski}},
  \bibinfo{author}{\bibfnamefont{M.}~\bibnamefont{Chollet}},
  \bibinfo{author}{\bibfnamefont{H.}~\bibnamefont{Lemke}},
  \bibinfo{author}{\bibfnamefont{R.}~\bibnamefont{Alonso-Mori}},
  \bibinfo{author}{\bibfnamefont{J.~M.} \bibnamefont{Glownia}},
  \bibinfo{author}{\bibfnamefont{J.}~\bibnamefont{Krzywinski}},
  \bibnamefont{et~al.}, \bibinfo{journal}{Proc. of SPIE}
  \textbf{\bibinfo{volume}{9210}}, \bibinfo{pages}{92100M}
  (\bibinfo{year}{2014}).

\bibitem[{\citenamefont{Goodman}(2000)}]{Goodman}
\bibinfo{author}{\bibfnamefont{J.~W.} \bibnamefont{Goodman}},
  \emph{\bibinfo{title}{Statistical Optics}} (\bibinfo{publisher}{Wiley},
  \bibinfo{address}{Hoboken, New Jersey}, \bibinfo{year}{2000}).

\bibitem[{\citenamefont{Mandel and Wolf}(1995)}]{MandelWolf}
\bibinfo{author}{\bibfnamefont{L.}~\bibnamefont{Mandel}} \bibnamefont{and}
  \bibinfo{author}{\bibfnamefont{E.}~\bibnamefont{Wolf}},
  \emph{\bibinfo{title}{Optical Coherence and Quantum Optics}}
  (\bibinfo{publisher}{Cambridge University Press},
  \bibinfo{address}{Cambridge, United Kingdom}, \bibinfo{year}{1995}).

\bibitem[{\citenamefont{Saldin et~al.}(2008)\citenamefont{Saldin,
  Schneidmiller, and Yurkov}}]{Saldin2008}
\bibinfo{author}{\bibfnamefont{E.~L.} \bibnamefont{Saldin}},
  \bibinfo{author}{\bibfnamefont{E.~A.} \bibnamefont{Schneidmiller}},
  \bibnamefont{and} \bibinfo{author}{\bibfnamefont{M.~V.}
  \bibnamefont{Yurkov}}, \bibinfo{journal}{Opt. Comm.}
  \textbf{\bibinfo{volume}{281}}, \bibinfo{pages}{1179} (\bibinfo{year}{2008}).

\bibitem[{\citenamefont{Vartanyants and Singer}(2010)}]{Vartanyants2010}
\bibinfo{author}{\bibfnamefont{I.~A.} \bibnamefont{Vartanyants}}
  \bibnamefont{and} \bibinfo{author}{\bibfnamefont{A.}~\bibnamefont{Singer}},
  \bibinfo{journal}{New J. Phys.} \textbf{\bibinfo{volume}{12}},
  \bibinfo{pages}{035004} (\bibinfo{year}{2010}).

\bibitem[{\citenamefont{Lajunen et~al.}(2005)\citenamefont{Lajunen, Vahimaa,
  and Tervo}}]{Lajunen2005}
\bibinfo{author}{\bibfnamefont{H.}~\bibnamefont{Lajunen}},
  \bibinfo{author}{\bibfnamefont{P.}~\bibnamefont{Vahimaa}}, \bibnamefont{and}
  \bibinfo{author}{\bibfnamefont{J.}~\bibnamefont{Tervo}}, \bibinfo{journal}{J.
  Opt. Soc. Am. A} \textbf{\bibinfo{volume}{22}}, \bibinfo{pages}{1536}
  (\bibinfo{year}{2005}).

\bibitem[{\citenamefont{Paul}(1986)}]{Paul1986}
\bibinfo{author}{\bibfnamefont{H.}~\bibnamefont{Paul}}, \bibinfo{journal}{Rev.
  Mod. Phys.} \textbf{\bibinfo{volume}{58}}, \bibinfo{pages}{209}
  (\bibinfo{year}{1986}).

\bibitem[{\citenamefont{Martins et~al.}(2006)\citenamefont{Martins, Wellhofer,
  Hoeft, Wurth, Feldhaus, and Follath}}]{Martins2006}
\bibinfo{author}{\bibfnamefont{M.}~\bibnamefont{Martins}},
  \bibinfo{author}{\bibfnamefont{M.}~\bibnamefont{Wellhofer}},
  \bibinfo{author}{\bibfnamefont{J.~T.} \bibnamefont{Hoeft}},
  \bibinfo{author}{\bibfnamefont{W.}~\bibnamefont{Wurth}},
  \bibinfo{author}{\bibfnamefont{J.}~\bibnamefont{Feldhaus}}, \bibnamefont{and}
  \bibinfo{author}{\bibfnamefont{R.}~\bibnamefont{Follath}},
  \bibinfo{journal}{Rev. Sci. Instrum.} \textbf{\bibinfo{volume}{77}},
  \bibinfo{pages}{115108} (\bibinfo{year}{2006}).

\bibitem[{\citenamefont{Gerasimova et~al.}(2011)\citenamefont{Gerasimova,
  Dziarzhytski, and Feldhaus}}]{Gerasimova2011}
\bibinfo{author}{\bibfnamefont{N.}~\bibnamefont{Gerasimova}},
  \bibinfo{author}{\bibfnamefont{S.}~\bibnamefont{Dziarzhytski}},
  \bibnamefont{and} \bibinfo{author}{\bibfnamefont{J.}~\bibnamefont{Feldhaus}},
  \bibinfo{journal}{J. Mod. Opt.} \textbf{\bibinfo{volume}{58}},
  \bibinfo{pages}{1480} (\bibinfo{year}{2011}).

\bibitem[{\citenamefont{Serkez}(2012)}]{Serkez2012}
\bibinfo{author}{\bibfnamefont{S.}~\bibnamefont{Serkez}}, Master's thesis,
  \bibinfo{school}{Ivan Franko Lviv National University and Deutsches
  Electronen Synchrotron}, \bibinfo{address}{Germany} (\bibinfo{year}{2012}).

\bibitem[{\citenamefont{Lutman et~al.}(2012)\citenamefont{Lutman, Ding, Feng,
  Huang, Messerschmidt, Wu, and Krzywinski}}]{Lutman2012}
\bibinfo{author}{\bibfnamefont{A.~A.} \bibnamefont{Lutman}},
  \bibinfo{author}{\bibfnamefont{Y.}~\bibnamefont{Ding}},
  \bibinfo{author}{\bibfnamefont{Y.}~\bibnamefont{Feng}},
  \bibinfo{author}{\bibfnamefont{Z.}~\bibnamefont{Huang}},
  \bibinfo{author}{\bibfnamefont{M.}~\bibnamefont{Messerschmidt}},
  \bibinfo{author}{\bibfnamefont{J.}~\bibnamefont{Wu}}, \bibnamefont{and}
  \bibinfo{author}{\bibfnamefont{J.}~\bibnamefont{Krzywinski}},
  \bibinfo{journal}{Phys. Rev. ST Accel. Beams} \textbf{\bibinfo{volume}{15}},
  \bibinfo{pages}{030705} (\bibinfo{year}{2012}).

\bibitem[{\citenamefont{Engel et~al.}(2016)\citenamefont{Engel, Dusterer,
  Brenner, and Teubner}}]{Engel2016}
\bibinfo{author}{\bibfnamefont{R.}~\bibnamefont{Engel}},
  \bibinfo{author}{\bibfnamefont{S.}~\bibnamefont{Dusterer}},
  \bibinfo{author}{\bibfnamefont{G.}~\bibnamefont{Brenner}}, \bibnamefont{and}
  \bibinfo{author}{\bibfnamefont{U.}~\bibnamefont{Teubner}},
  \bibinfo{journal}{J. Synch. Rad.} \textbf{\bibinfo{volume}{23}},
  \bibinfo{pages}{118} (\bibinfo{year}{2016}).

\bibitem[{\citenamefont{Saldin et~al.}(1998)\citenamefont{Saldin,
  Schneidmiller, and Yurkov}}]{Saldin1998}
\bibinfo{author}{\bibfnamefont{E.~L.} \bibnamefont{Saldin}},
  \bibinfo{author}{\bibfnamefont{E.~A.} \bibnamefont{Schneidmiller}},
  \bibnamefont{and} \bibinfo{author}{\bibfnamefont{M.~V.}
  \bibnamefont{Yurkov}}, \bibinfo{journal}{Opt. Comm.}
  \textbf{\bibinfo{volume}{148}}, \bibinfo{pages}{383 } (\bibinfo{year}{1998}).

\bibitem[{\citenamefont{Engel}(2015)}]{Engel2015}
\bibinfo{author}{\bibfnamefont{R.}~\bibnamefont{Engel}},
  \bibinfo{type}{Bachelor's thesis}, \bibinfo{school}{Carl-von-Ossietzky
  Universitat at Oldenburg and Deutsches Electronen Synchrotron},
  \bibinfo{address}{Germany} (\bibinfo{year}{2015}).

\bibitem[{\citenamefont{Breitenbach et~al.}(1997)\citenamefont{Breitenbach,
  Schiller, and Mlynek}}]{Breitenbach1997}
\bibinfo{author}{\bibfnamefont{G.}~\bibnamefont{Breitenbach}},
  \bibinfo{author}{\bibfnamefont{S.}~\bibnamefont{Schiller}}, \bibnamefont{and}
  \bibinfo{author}{\bibfnamefont{J.}~\bibnamefont{Mlynek}},
  \bibinfo{journal}{Nature} \textbf{\bibinfo{volume}{387}}, \bibinfo{pages}{471
  } (\bibinfo{year}{1997}).

\bibitem[{\citenamefont{Oppel et~al.}(2012)\citenamefont{Oppel, B\"uttner, Kok,
  and von Zanthier}}]{Oppel2012}
\bibinfo{author}{\bibfnamefont{S.}~\bibnamefont{Oppel}},
  \bibinfo{author}{\bibfnamefont{T.}~\bibnamefont{B\"uttner}},
  \bibinfo{author}{\bibfnamefont{P.}~\bibnamefont{Kok}}, \bibnamefont{and}
  \bibinfo{author}{\bibfnamefont{J.}~\bibnamefont{von Zanthier}},
  \bibinfo{journal}{Phys. Rev. Lett.} \textbf{\bibinfo{volume}{109}},
  \bibinfo{pages}{233603} (\bibinfo{year}{2012}).

\bibitem[{\citenamefont{Pelliccia et~al.}(2016)\citenamefont{Pelliccia, Rack,
  Scheel, Cantelli, and Paganin}}]{Pelliccia2016}
\bibinfo{author}{\bibfnamefont{D.}~\bibnamefont{Pelliccia}},
  \bibinfo{author}{\bibfnamefont{A.}~\bibnamefont{Rack}},
  \bibinfo{author}{\bibfnamefont{M.}~\bibnamefont{Scheel}},
  \bibinfo{author}{\bibfnamefont{V.}~\bibnamefont{Cantelli}}, \bibnamefont{and}
  \bibinfo{author}{\bibfnamefont{D.~M.} \bibnamefont{Paganin}},
  \bibinfo{journal}{Phys. Rev. Lett.} \textbf{\bibinfo{volume}{117}},
  \bibinfo{pages}{113902} (\bibinfo{year}{2016}).

\bibitem[{\citenamefont{Yu et~al.}(2016)\citenamefont{Yu, Lu, Han, Xie, Du,
  Xiao, and Zhu}}]{Yu2016}
\bibinfo{author}{\bibfnamefont{H.}~\bibnamefont{Yu}},
  \bibinfo{author}{\bibfnamefont{R.}~\bibnamefont{Lu}},
  \bibinfo{author}{\bibfnamefont{S.}~\bibnamefont{Han}},
  \bibinfo{author}{\bibfnamefont{H.}~\bibnamefont{Xie}},
  \bibinfo{author}{\bibfnamefont{G.}~\bibnamefont{Du}},
  \bibinfo{author}{\bibfnamefont{T.}~\bibnamefont{Xiao}}, \bibnamefont{and}
  \bibinfo{author}{\bibfnamefont{D.}~\bibnamefont{Zhu}},
  \bibinfo{journal}{Phys. Rev. Lett.} \textbf{\bibinfo{volume}{117}},
  \bibinfo{pages}{113901} (\bibinfo{year}{2016}).

\bibitem[{\citenamefont{Prince et~al.}(2016)\citenamefont{Prince, Allaria,
  Callegari, Cucini, De~Ninno, Di~Mitri, Diviacco, Ferrari, Finetti, Gauthier
  et~al.}}]{Prince2016}
\bibinfo{author}{\bibfnamefont{K.~C.} \bibnamefont{Prince}},
  \bibinfo{author}{\bibfnamefont{E.}~\bibnamefont{Allaria}},
  \bibinfo{author}{\bibfnamefont{C.}~\bibnamefont{Callegari}},
  \bibinfo{author}{\bibfnamefont{R.}~\bibnamefont{Cucini}},
  \bibinfo{author}{\bibfnamefont{G.}~\bibnamefont{De~Ninno}},
  \bibinfo{author}{\bibfnamefont{S.}~\bibnamefont{Di~Mitri}},
  \bibinfo{author}{\bibfnamefont{B.}~\bibnamefont{Diviacco}},
  \bibinfo{author}{\bibfnamefont{E.}~\bibnamefont{Ferrari}},
  \bibinfo{author}{\bibfnamefont{P.}~\bibnamefont{Finetti}},
  \bibinfo{author}{\bibfnamefont{D.}~\bibnamefont{Gauthier}},
  \bibnamefont{et~al.}, \bibinfo{journal}{Nat. Photon.}
  \textbf{\bibinfo{volume}{10}}, \bibinfo{pages}{176} (\bibinfo{year}{2016}).

\bibitem[{\citenamefont{Usenko et~al.}(2016)\citenamefont{Usenko, Przystawik,
  Jakob, Lazzarino, Brenner, Toleikis, Ch.~Haunhorst, and
  Laarmann}}]{Usenko2016}
\bibinfo{author}{\bibfnamefont{S.}~\bibnamefont{Usenko}},
  \bibinfo{author}{\bibfnamefont{A.}~\bibnamefont{Przystawik}},
  \bibinfo{author}{\bibfnamefont{M.}~\bibnamefont{Jakob}},
  \bibinfo{author}{\bibfnamefont{L.~L.} \bibnamefont{Lazzarino}},
  \bibinfo{author}{\bibfnamefont{G.}~\bibnamefont{Brenner}},
  \bibinfo{author}{\bibfnamefont{S.}~\bibnamefont{Toleikis}},
  \bibinfo{author}{\bibfnamefont{D.~K.} \bibnamefont{Ch.~Haunhorst}},
  \bibnamefont{and} \bibinfo{author}{\bibfnamefont{T.}~\bibnamefont{Laarmann}},
  \bibinfo{journal}{Nature Commun.}  (\bibinfo{year}{2016}),
  \bibinfo{note}{submitted}.

\bibitem[{\citenamefont{Bracewell}(1999)}]{FourierBook}
\bibinfo{author}{\bibfnamefont{R.}~\bibnamefont{Bracewell}},
  \emph{\bibinfo{title}{The Fourier Transform and Its Applications, 3rd ed}}
  (\bibinfo{publisher}{McGraw-Hill}, \bibinfo{address}{New York},
  \bibinfo{year}{1999}).

\bibitem[{\citenamefont{Weisstein}()}]{convolution}
\bibinfo{author}{\bibfnamefont{E.~W.} \bibnamefont{Weisstein}},
  \emph{\bibinfo{title}{Convolution. {From MathWorld---A Wolfram Web
  Resource}}},
  \urlprefix\url{\url{http://mathworld.wolfram.com/Convolution.html}}.

\end{thebibliography}

\eject

\begin{table}
        \caption{\label{tab::sim_param} Beam parameters used in simulations}
        \begin{center}
                \begin{tabularx}{\textwidth}{l
 >{\centering\let\newline\\\arraybackslash\hspace{0pt}}X|>{\centering\let\newline\\\arraybackslash\hspace{0pt}}X >{\centering\let\newline\\\arraybackslash\hspace{0pt}}X |>{\centering\let\newline\\\arraybackslash\hspace{0pt}}X>{\centering\let\newline\\\arraybackslash\hspace{0pt}}X}
                        \hline
                        \hline
                        Simulation                 & \multicolumn{1}{c}{Model I}   & \multicolumn{2}{c}{Model II}    & \multicolumn{2}{c}{Model III} \\
                        \hline
                        Beam number                    &   1  & 1      & 2         & 1    &   2   \\
                        \hline
                        Bandwidth, $D_{\omega}/\omega$ & \multicolumn{1}{c}{$10^{-4}$}   & \multicolumn{2}{c}{$10^{-4}$}    & \multicolumn{2}{c}{$10^{-4}$} \\
                        Relative intensity             &  1    & 1      & 0.01         & 1        &   0.1   \\
                        Beam position, $x_0$ (mm)                                                &        0  & 0                 & 0.6      & 0                & -0.5   \\
                        Beam size (rms), $\sigma_x$ (mm)               &  0.7  & 0.4       & 0.5        & 0.4          & 0.4       \\
                        Transverse coherence length, $\xi_x$ (mm)                   & 1   & 0.6         & 3        & 0.8         & 0.8       \\
                        Central frequency, $\omega_0$ (fs$^{-1}$)        & 140  & 140                  & 140             & 140           & 140     \\
                        Spectral width, $\Omega$ (fs$^{-1}$)        & 0.01  & 0.01         & 0.01           & 0.01          & 0.01  \\
                        Spectral coherence width, $\Omega_c$ (fs$^{-1}$)         & 0.007  & 0.0014           & 0.0028         & 0.008                 & 0.014   \\
                        \hline
                        \hline
                \end{tabularx}
        \end{center}
\end{table}

\begin{table}
	\caption{\label{tab::parameters} FLASH and PG2 beamline parameters}
	\begin{center}
	\begin{tabularx}{\textwidth}{l >{\centering\let\newline\\\arraybackslash\hspace{0pt}}X >{\centering\let\newline\\\arraybackslash\hspace{0pt}}X >{\centering\let\newline\\\arraybackslash\hspace{0pt}}X}
		\hline
		\hline
		Wavelength (nm)                               & 5.5           & 13.4     & 20.8  	\\
		\hline
		Photon energy (eV) 							  & 226 		  & 94       & 59 		\\
		Charge (nC)                               	  & 0.6           & 0.3     & 0.2   	\\
		Pulse energy ($\mu$J)                        & 110           & 21    & 30-40 	\\
		Grating order 								  & 3  			  & 1 		 & 2 		\\
		Grating parameter $c_{ff}$ 					 & 2 			  & 1.5 	 & 1.7 		\\
		Dispersion (eV/mm) 						  & 0.64 		  & 0.35 	 & 0.11 	\\
		Focus to detector distance (m) 				  & 3.3 		  & 1.6 	 & 2.5 		\\
		\hline
		\hline
	\end{tabularx}
	\end{center}
\end{table}

\begin{table}
	\caption{\label{tab::results} Results of HBT measurements}
	\begin{center}
	\begin{tabularx}{\textwidth}{l >{\centering\let\newline\\\arraybackslash\hspace{0pt}}X >{\centering\let\newline\\\arraybackslash\hspace{0pt}}X >{\centering\let\newline\\\arraybackslash\hspace{0pt}}X}
		\hline
		\hline
		Wavelength (nm)                                          & 5.5           & 13.4     & 20.8  	\\
		\hline
		Average pulse size (FWHM) ($\mu$m)                       & 450           & 220      & 780     \\
		Coherence length $l_c$ ($\mu$m)                          & 930           & 240      & 590  \\
		Degree of spatial coherence $\zeta_s$                    & 78\%          & 85\%     & 48\% \\
		Pulse duration $T$ (fs), from HBT measurements           & 31            & 61       & 21   \\
		Pulse duration $T$ (fs), from spectral measurements      & 27            & 55       & 27 \\
		\hline
		\hline
	\end{tabularx}
	\end{center}
\end{table}

\printtables

\eject

\end{document}